\documentclass{article}

\pdfoutput=1

\usepackage{latexsym,amsfonts,amsmath,url,bm} 
\usepackage{boxedminipage,graphicx,color,array,multicol,subfigure}
\usepackage[abs]{overpic}
\usepackage{amsthm}
\usepackage[utf8]{inputenc}
\usepackage[english]{babel}
\usepackage{fancyhdr}

\usepackage[utf8]{inputenc}
\usepackage{graphicx}
\usepackage{authblk}
\usepackage[colorlinks=true,citecolor=blue]{hyperref}
\usepackage{url}
\usepackage{csquotes}
\usepackage{amsmath}
\usepackage{amssymb}
\usepackage[dvipsnames]{xcolor}
\usepackage{mathtools}
\DeclarePairedDelimiter{\ceil}{\lceil}{\rceil}
\usepackage{footmisc}
\usepackage{caption}

\newcommand{\eg}{e.g.~}
\newcommand{\ie}{i.e.~}

\newcommand{\cf}{cf.~}
\newcommand{\uscb}{U.S.\ Census Bureau}
\newcommand{\lquote}[1]{``{\it#1}''}

\newcommand{\dist}{\mathbf{dist}}
\newcommand{\expbf}{\mathbf{exp}}
\newcommand{\poly}{\mathbf{poly}}
\newcommand{\order}{\mathcal{O}}
\newcommand{\lap}{\mathrm{Lap}}
\newcommand{\var}{\mathrm{Var}}
\newcommand{\pr}{\mathrm{Pr}}
\newcommand{\eps}{\varepsilon}
\newcommand{\epsdel}{(\eps,\delta)}
\newcommand{\percent}{\,\%}
\newcommand{\define}{\coloneqq}
\newcommand{\ptable}{$p$-table}
\newcommand{\tot}{\textrm{total}}

\title{
    Differential privacy and noisy confidentiality concepts for European population statistics
    }

\author[1]{Fabian Bach
    \thanks{
    \href{mailto:Fabian.BACH@ec.europa.eu}{Fabian.BACH@ec.europa.eu}}
    }

\affil[1]{European Commission, Eurostat, L-2920 Luxembourg\thanks{
    The views expressed are purely those of the authors and may not in any circumstances be regarded as stating an official position of the European Commission.}
    }

\date{\today}

\begin{document}

\maketitle

\begin{abstract}
    The paper aims to give an overview of various approaches to statistical disclosure control based on random noise that are currently being discussed for official population statistics and censuses.  A particular focus is on a stringent delineation between different concepts influencing the discussion: we separate clearly between risk measures, noise distributions and output mechanisms---putting these concepts into scope and into relation with each other.
    
    After recapitulating differential privacy as a risk measure, the paper also remarks on utility and risk aspects of some specific output mechanisms and parameter setups, with special attention on static outputs that are rather typical in official population statistics.  In particular, it is argued that unbounded noise distributions, such as plain Laplace, may jeopardise key unique census features without a clear need from a risk perspective.  On the other hand, bounded noise distributions, such as the truncated Laplace or the cell key method, can be set up to keep unique census features while controlling disclosure risks in census-like outputs.
    
    Finally, the paper analyses some typical attack scenarios to constrain generic noise parameter ranges that suggest a good risk/utility compromise for the 2021 EU census output scenario.  The analysis also shows that strictly differentially private mechanisms would be severely constrained in this scenario.
\end{abstract}

\section{Background}\label{intro}

After Dinur and Nissim published their seminal database reconstruction theorem almost two decades ago~\cite{dinur2003}, it has shaped and accelerated research activities across many domains involved with data protection, data privacy and confidentiality, including disclosure control in official statistics.  In its wake, `differential privacy' was proposed in 2006~\cite{dwork2006,dwork2006a} initially as a rigorous privacy or risk measure addressing consequences from the database reconstruction theorem.  Differentially private noise mechanisms were then picked up and developed further to test and improve its use for (official) statistics; see \eg~\cite{machanavajjhala2008,hardt2010,ghosh2012,dwork2014,dwork2016,rinott2018}.

Now a first strict line must be drawn between differential privacy as a {\it risk measure}, and differentially private (noisy) {\it output mechanisms} that are engineered to manifestly guarantee a given differential privacy level.  However, many other noisy output mechanisms, using bounded or unbounded noise distributions, can be set up to give at least a relaxed differential privacy guarantee too~\cite{dwork2006a,rinott2018}.  For instance, the cell key method originally proposed by the Australian Bureau of Statistics~\cite{fraser2005,marley2011,thompson2013} can be turned into a (relaxed) differentially private mechanism~\cite{bailie2019}.  On the other hand, strictly differentially private output mechanisms require unbounded noise distributions with infinite tails, which may have particularly negative effects on utility.  This paper aims to first address all these different notions separately, and then to present a consolidated discussion from both risk and utility perspectives.

We further focus on population and census-like statistics with typical outputs being unweighted person counts possibly arranged in contingency tables.  This serves
two distinct motivations: On the one hand, treating only unweighted counts
simplifies many technicalities without touching key issues of the noise discussion.  On the other hand, global efforts on the 2020/2021 census round are peaking right now, with many important (and urgent) contact points to this paper.  For instance, the \uscb\ has adopted a strictly differentially private noise mechanism for the 2020 U.S.\ census~\cite{abowd2018,garfinkel2019,petti2019}, which received mixed reactions down to grave utility concerns~\cite{ruggles2019,santos2020}.  On the other hand, the European Statistical System\footnote{The joint body of Eurostat and the national statistical institutes of all EU countries and Iceland, Liechtenstein, Norway and Switzerland.  It is responsible for the development and quality assurance of official European statistics.\label{fn_ess}}
has developed recommendations for a harmonised protection of 2021 EU census outputs based on the cell key method~\cite{essnet2017,essnet2019,essnet2019a}, where sizeable disclosure risks from massive averaging attacks were claimed recently~\cite{ashgar2019}.  Also these issues will be put into scope in the further course.

Our goals are to give a comparative overview of the various terms and concepts, and to present some analytic evidence that may contribute to the process of setting up an appropriate noise mechanism for particular output scenarios of official population or census statistics.  The paper is structured as follows: section~\ref{pre} prepares the stage by introducing terminology and outlining the different concepts separately; section~\ref{risk} addresses context relevant risk issues, such as database reconstruction and typical attacks on census outputs including re-identification; section~\ref{utility} introduces some specific noise setups currently discussed in a census context, and elaborates on utility aspects focusing on unique census features; and section~\ref{censuses} finally offers a brief synthesis in view of ongoing census efforts including some evidence-based generic constraints on noise setups before the paper concludes.

\section{Preliminaries and concepts}\label{pre}

\subsection{Database reconstruction theorem}\label{pre_db_recon}

In their 2003 paper~\cite{dinur2003}, Dinur and Nissim have shown on very generic grounds that input databases to a query system returning counts with bounded noise of magnitude $\leq E$ can be reconstructed to a high accuracy, if a large enough number of independent query results is available (called ``output complexity'').  More explicitly, Theorem~2 proves that the Hamming distance between the true input database $d$ containing $n$ Boolean records ($n$ entries of $0$ or $1$) and the reconstructed candidate $c$ is
\begin{equation*}
    \dist\left( d,c \right) \leq 4E,
\end{equation*}
if the output complexity is an exponential in $n$ (called ``$\expbf(n)$-non-privacy'').  This means noise magnitude up to the order of $n$, or $E\lesssim\order(n)$, would be needed to provide effective protection, thus swamping any output utility.\footnote{In this paper we employ a slack notation using `$\order(X)$' to be read as `of the order (of magnitude) of $X$' without further assumptions on $X$.  The typeset $\order$ is to avoid confusion with more rigid (and heavier) limiting behaviour notations that typically use $O$, $o$, $\Omega$, etc.\label{fn_order_notation}}
However, $\expbf(n)$ complexity of the output is obviously an extreme scenario with marginal practical relevance, so that the following theorem is the one that became famous as the \lquote{data base reconstruction theorem}, or sometimes \lquote{fundamental law of information recovery}:

Theorem~3 of~\cite{dinur2003} proves that an upper limit can also be placed on the inaccuracy of a {\it polynomial} reconstruction attack, \ie the output complexity is at most a power of~$n$ (fixed here to $t=n\log^2 n$), if the noise magnitude is $E<\order(\sqrt{n})$ (called ``$\poly(n)$-non-privacy'').\footnote{Note that in the algorithm these queries are random, so not tailored by the attacker.  This means also static output tables with $\poly(n)$ cells fall under the theorem in principle.\label{poly}}
This upper limit is
\begin{equation}\label{eq_poly-np}
    \dist\left( d,c \right) \leq \epsilon n ,
\end{equation}
where $\epsilon<1$ is an arbitrary accuracy parameter.  The proof makes no assumption on the type of noise, except that its magnitude is strictly bounded by $E<\order(\sqrt{n})$.  Therefore, any noise distribution with moderate (strictly) bounded noise is susceptible in principle, if the output has $\poly(n)$ complexity.

This initially suggests two options: do not bound the noise or limit the output complexity; we will come back to both in section~\ref{risk}.  In any case, Theorem~5 of~\cite{dinur2003} already shows that provably private query systems can be obtained through noise mechanisms with magnitude scaling as $\order(\sqrt{t})$, where $t$ is a given output complexity of the query system; \cf further investigations in this direction, \eg \cite{dwork2004,blum2005}.  Apart from the reconstruction theorem itself, this is the second important conclusion from~\cite{dinur2003}: there is a scaling law requiring that noise variance should increase with output complexity, or inversely, the noise variance can be fixed to a sufficient constant if the output complexity is fixed too.

\subsection{Differential privacy: a risk measure}\label{pre_dp}

Differential privacy was first proposed in 2006~\cite{dwork2006}, in the wake of the database reconstruction theorem.  In plain words, its paradigm is that every query result (output) should be robust against addition to, or removal from, the input database of any single record, \eg picking one record and removing it from the database should not significantly change any outputs (hence {\it differential} privacy).  This is the individual privacy guarantee, and its immediate attraction is formulated in~\cite{dwork2011}: \lquote{Any mechanism satisfying this definition addresses all concerns that any participant might have about the leakage of his or her personal information, regardless of any auxiliary information known to an adversary: Even if the participant removed his or her data from the dataset, no outputs (and thus consequences of outputs) would become significantly more or less likely.}

There are various mathematical definitions of differential privacy, so we repeat here the most generic one, introducing both strict as well as relaxed (or approximate) differential privacy in one go~\cite{dwork2006a}: given two neighbouring input databases $d$ and $d^\prime$ that differ exactly in one record, any mechanism $\mathcal{M}(\cdot)$ acting on the universe of input databases to generate outputs must fulfil
\begin{equation}\label{eq_def_dp_epsdel}
    \pr(\mathcal{M}(d)\in S) \leq e^\eps\pr(\mathcal{M}(d^\prime)\in S) + \delta
\end{equation}
for all subsets $S\subseteq \mathrm{Range}(\mathcal{M})$ to be {\it $\delta$-approximately $\eps$-differentially private} or short $\epsdel$-DP, where $\eps$ and $\delta$ are parameters establishing the differential privacy level.  For $\delta\rightarrow 0$, Eq.~\eqref{eq_def_dp_epsdel} reduces to a definition of {\it strictly $\eps$-differentially private} or short $\eps$-DP mechanisms.

The definition implies that, for any single output $s\in \mathrm{Range}(\mathcal{M})$---singleton $S$ in Eq.~\eqref{eq_def_dp_epsdel}---with nonzero probability on $d$, the probability to obtain $s$ from $d^\prime$ should also be nonzero for the mechanism to be possibly $\eps$-DP.  This suggests some kind of noise injection applied by $\mathcal{M}$ as an option to comply with Eq.~\eqref{eq_def_dp_epsdel}.  While noisy $\mathcal{M}$s are discussed in more detail in section~\ref{pre_output}, it is important to note here that $\eps$-DP or $\epsdel$-DP are attributes or qualifiers of any given $\mathcal{M}$, thus measuring the individual information leakage from any thinkable output. Therefore, $\eps$ and $\delta$ are handy risk measures to compare different output scenarios and noise mechanisms, as done \eg in~\cite{rinott2018}.

Finally, it is interesting to note how differential privacy embraces the {\it fundamental law of statistical disclosure control} (formalised  in~\cite{dwork2010} as {\it rigorous impossibility of Dalenius's privacy goal}), which essentially states that {\it any} provision of useful statistical information {\it necessarily} entails a nonzero trailing risk of disclosing information on some individuals\footnote{Interestingly, including potentially individuals that did not even contribute to the statistics, as pointed out \eg in~\cite{dwork2011}.},
\ie that the trade-off between privacy and utility is fundamental and not contingent.  With the privacy budget parameter~$\eps$, differential privacy provides a transparent and intuitive ``turning knob''---but no immediate guidance on how exactly to adjust it.

\subsection{Noise distributions: bounded or unbounded?}\label{pre_noise}

Recall that the discussion is confined to outputs representing unweighted person counts, or sets of such counts (\eg contingency tables).
Then the most generic output mechanism $\mathcal{M}(\cdot)$, in the sense of sections~\ref{pre_dp} and~\ref{pre_output}, returns an ordered $k$-tuple of frequencies representing the answers to $k$ individual counting queries passed to $\mathcal{M}$.
Further let $\widetilde{\mathcal{M}}(\cdot)$ denote an {\it exact} output mechanism without any noise injected, so that $\mathrm{Range}(\widetilde{\mathcal{M}})=\mathbb{N}_0^k$.  Then by noise distribution we mean the probability distribution underlying the process of drawing an additive (pseudo) random noise term $x\equiv(\mathcal{M}-\widetilde{\mathcal{M}})(d)$ for $k=1$ and any given $d$.  Among the popular options are \eg Laplace, Gaussian, or entropy-maximising distributions, which may come in various flavours and with auxiliary constraints, but many properties can be captured by just two generic attributes: the noise variance $\var(x)$ and its magnitude bound $\left|x\right|\leq E$ ($\leq\infty$).  Here we just give a crude classification based on the DP categories introduced in section~\ref{pre_dp}.

\textbf{$\eps$-DP noise distributions}\quad manifestly comply with Eq.~\eqref{eq_def_dp_epsdel} for any possible singleton $S$ (single output count) with $\delta=0$.  It is easy to show~\cite{dwork2011} that \eg the Laplace distribution
\begin{equation}\label{eq_lap}
    \lap\left(\Delta/\eps\right):\; x\sim \frac{\eps}{2\Delta}\exp\left(-\frac{\eps|x|}{\Delta}\right)
\end{equation}
with $\var(x)=2(\Delta/\eps)^2$ fulfils this requirement, where $\Delta$ is the {\it global sensitivity} of $\widetilde{\mathcal{M}}$ defined as
\begin{equation}\label{eq_df}
    \Delta \define \max_{d,d^\prime} \sum_{i=1}^k \left| \widetilde{\mathcal{M}}(d)_i - \widetilde{\mathcal{M}}(d^\prime)_i \right|
\end{equation}
with $i$ running through output $k$-tuple indices.  Clearly for $k=1$ and unweighted person counts, $\Delta=1$ and $x\sim\lap(1/\eps)$ in this case.\footnote{Apart from unweighted counts, the issue with $\Delta$ is that it is generally hard to obtain, and arbitrarily difficult for some queries on weighted or magnitude data: for instance~\cite{bambauer2014}, in an average income query the global sensitivity is theoretically driven by the highest-income person in the world, because the query result must be robust also against addition of that person to the database.  Naturally such a $\Delta$ drives the noise through the roof and renders all outputs useless.  On the other hand, capping $\Delta$ arbitrarily dilutes the individual privacy guarantee.\label{fn_df}}
Now this distribution is over $\mathbb{R}$, so that $\mathrm{Range}(\mathcal{M})=\mathbb{R}^k$ which may return non-integer person counts.  This can be lifted by using the discrete two-tailed geometric distribution~\cite{ghosh2012}
\begin{equation}\label{eq_geom2}
    x\sim \frac{1-\exp(-\eps)}{1+\exp(-\eps)} \exp(-\eps|x|),
\end{equation}
which gives $\mathrm{Range}(\mathcal{M})=\mathbb{Z}^k$ and approximates to $\lap(1/\eps)$ for $\eps\ll 1$.

Note finally that noise distributions, continuous or discrete, must be unbounded to be $\eps$-DP.  To see this, assume bounded noise with $\pr(x>E)=0$.  Then in Eq.~\eqref{eq_def_dp_epsdel}, choose without loss of generality $d>d^\prime$ and $s=\widetilde{\mathcal{M}}(d)+E$ (\ie $k=1$).  Thus, $\pr(\mathcal{M}(d^\prime)=s)=0$ as $s-\widetilde{\mathcal{M}}(d^\prime)=E+1$ and the inequality requires $\delta\geq\pr(\mathcal{M}(d)=s)>0$ to hold, which contradicts $\delta=0$.

\textbf{$\epsdel$-DP and other noise distributions}\quad In a sloppy manner, most noise distributions that are not $\eps$-DP are $\epsdel$-DP: If a distribution fails the strict $\eps$-DP requirement, Eq.~\eqref{eq_def_dp_epsdel} with $\delta=0$, a $\delta>0$ can usually be found to establish $\epsdel$-DP.  In particular, unbounded noise distributions can usually be truncated to give $\epsdel$-DP, where $\delta$ depends on the resulting probability distribution close to its discontinuity~\cite{rinott2018}.  Also for cell key noise~\cite{thompson2013}, taking variance $\var(x)\equiv V$ and noise bound $|x|\leq E$ as input parameters, an $\epsdel$-DP level can be inferred~\cite{bailie2019}.  However, the issue is not about finding a $\delta$ but about dealing with its value: clearly it should be $\delta\ll 1$ but how small exactly? For instance, $\delta<1/n$ is stated in~\cite{dwork2014}, but higher values are also discussed in~\cite{rinott2018}.  It is also argued there that often the choices of $\delta$ (and $\eps$) are policy decisions, not statistical decisions.

\textbf{Bounded vs.\ unbounded noise}\quad  Why select an unbounded noise distribution?  As shown above, if a strictly $\eps$-DP output mechanism is ultimately desired, the underlying noise distribution must be unbounded.  Moreover, it was recently claimed that a tight noise bound poses additional disclosure risks~\cite{ashgar2019}.  However, sections~\ref{risk} and \ref{utility} will argue that unbounded noise may come at too high a price on utility, while the additional risks of bounded noise can be controlled.\footnote{All proofs in~\cite{dinur2003} relied on a tight noise bound, which may lead to the impression that unbounded noise a priori avoids the premises of the database reconstruction theorem; this point will be touched in section~\ref{risk_DP}.\label{fn_eps-DP_scaling}}

\subsection{Noisy output mechanisms}\label{pre_output}

Noise distributions handle the special case of a single scalar output, \ie a single call to $\mathcal{M}$ with $k=1$.  In contrast, a generic (noisy) output mechanism denotes a more powerful and complex $\mathcal{M}$ that ideally accounts automatically for {\it noise composition} across all outputs.  In particular, from a DP perspective an $\eps$-DP noise distribution does not automatically constitute an $\eps^\prime$-DP output mechanism with $\eps=\eps^\prime$; as argued below, an additional layer of output curation or privacy budget management is needed.  But before outlining various output mechanisms, some prerequisites are introduced.

\textbf{Disjoint outputs: histograms and tables}\quad are lists of counts breaking down the input database into sub-populations, \ie (using notation introduced in annex~\ref{a_av_output}) a single call to an $\mathcal{M}:\,d\mapsto T_A$ where $A=\{a_i\}$ is a list of disjoint breakdown categories contained in $d$ and obviously $k=|A|$.  It is easy to see from Eq.~\eqref{eq_df} that still $\Delta=1$ despite $k>1$ output counts, because addition or removal of a single record in $d$ can only change the single count $\mathcal{M}(d)_i\equiv T_{a_i}$ where that record contributes.\footnote{Note that we decided in annex~\ref{a_av_output} to suppress all categories $a_i=\mathrm{total}$; if the total count was included in $A$, $\Delta=2$.}
In DP literature, this class of queries is called histogram queries (\eg \cite{dwork2011}), and properties transcend to multi-dimensional table outputs $\mathcal{M}:\,d\mapsto T_\mathbf{A}$ with $\mathbf{A}=\{A_i\}$ and $k=\prod_{i=1}^{|\mathbf{A}|}|A_i|$ but $\Delta=1$ still.

\textbf{Same participants--same noise (SPSN)}\quad is a principle to decide whether the noise term added to a given output count is drawn afresh or reused from a lookup table~\cite{fraser2005,thompson2013}.  It was introduced to forestall averaging to some extent, as per Chebyshev's inequality the probability that $t$ redundant noisy observations average to the true count converges to 1 for increasing $t$ (\cf Eq.~\eqref{eq_cheb}).  If the noise is looked up instead to be always the same for the same question asked, averaging over redundancies is less straightforward (but still possible, as shown in section~\ref{risk_BN}).\footnote{The intricacies of defining SPSN discussed in~\cite{rinott2018} are not relevant here, because noise independence is established contextually through the variable attributes $\mathbf{a}$ of a given cell count $T_\mathbf{a}$.\label{fn_same_noise}}

\textbf{Static output mechanisms: non-interactive $\mathcal{M}$}\quad With notions of table outputs and SPSN at hand, a static output mechanism without SPSN is defined here as
\begin{equation}\label{eq_M-static_no_SPSN}
    \mathcal{M}:\,d\mapsto \cup_{I\in\{1\cdots M\}} TI_{\mathcal{P}(\mathbf{A}_I)}
\end{equation}
returning $M$ tables and all marginals in a single call ($\mathcal{P}(\mathbf{A})$ is the power set of $\mathbf{A}$).  Think of $\mathcal{M}$ returning the entire set of $M=103$ population tables in the 2021 EU census programme\footnote{Commission Regulation (EU) 2017/712 of 20 April 2017 establishing the reference year and the programme of the statistical data and metadata for population and housing censuses provided for by Regulation (EC) No 763/2008 of the European Parliament and of the Council (\href{https://eur-lex.europa.eu/legal-content/EN/TXT/?qid=1493361675727&uri=CELEX:32017R0712}{OJ~L~105, 21.4.2017, p.~1}).\label{fn_cir-2}}
at once. If SPSN is invoked,
\begin{equation}\label{eq_M-static_SPSN}
    \mathcal{M}:\,d\mapsto \cup_{I\in\{1\cdots M\}} T_{\mathcal{P}(\mathbf{A}_I)}
\end{equation}
containing only unique cross-tabulations in the table programme, which is a much smaller set than in Eq.~\eqref{eq_M-static_no_SPSN}.  Then the $M$ predefined tables can be put together for publication from the unique outputs (\eg collecting internal cells and all corresponding marginals).  This also shows how a static output mechanism can be combined with a more flexible user interface (table builder):  Just expose the unique outputs as building blocks available for user-defined tables.

The advantage of a static output is that the curator has full control over the output and so decides about the acceptable amount of redundancies and internal constraints (the levers for typical disclosure attacks described in section~\ref{risk}) before publication.  Moreover, $k$ and $\Delta$ for a DP mechanism can simply be counted from the redundancies in the output set; the full exercise is carried out in annex~\ref{a_av}, and $k$ and $\Delta$ are given by Eq.~\eqref{eq_k_t_global} (without SPSN) resp.\ Eq.~\eqref{eq_k_t_unique} (with SPSN).

\textbf{Flexible output mechanisms: interactive $\mathcal{M}$}\quad It is sometimes argued (justly) that static outputs cannot provide the full richness of $d$, at least not without running into the database reconstruction theorem (\eg \cite{dwork2006,dwork2011}).  For the scope of this paper, a flexible output mechanism is again defined as $\mathcal{M}:\,d\mapsto T_\mathbf{A}$, but this time with interactive elements: the cross-classification $\mathbf{A}=\{A_i\}$ as well as variable breakdowns $A_i=\{a_{ij}\}$ are (at least to some extent) customisable, and the user is allowed to call $\mathcal{M}$ $t$ times with a series of questions $\{\mathbf{A}_t\}$.  For each $\mathbf{A}_t$ the noise can be curated but a priori, $t$ is unknown.  Such an output mechanism without any noise curation across $t$ is highly susceptible to various attacks described in~\cite{ashgar2019}, and discussed again in section~\ref{risk_av}.  DP may guide a way out, but at a costly price, as shown below.  Anyway it seems that, in certain scenarios such as censuses, static outputs are preferable because the objectives are clear and limited, so that outputs can be curated accordingly.

\textbf{Manifestly $\eps$-DP output mechanisms}\quad In static outputs with an $\eps$-DP noise distribution, the global privacy budget $\eps$ is automatically distributed correctly across all outputs by virtue of $\Delta$.  Customised distributions of $\eps$ are possible, using \eg $i$ calls to $\mathcal{M}$ with $\eps_i$ each.  Then the whole output is still $\eps$-DP with $\eps=\sum_i\eps_i$ through the DP composition theorem~\cite{dwork2014,rinott2018}.  This is still a static mechanism, and it is what the \uscb\ plans for its 2020 census outputs~\cite{abowd2018,garfinkel2019}.

Flexible mechanisms can also be $\eps$-DP through composition but are more tricky, because ad hoc rules must be applied to distribute a global $\eps$ across $t$ outputs.  Two approaches are usually proposed~\cite{dwork2011}: either cap $t$ per user at some value, or spend $\eps$ iteratively as $\eps_i=\eps/2^i$, so that $\eps=\sum_{i=1}^{\infty}
\eps_i$.  Both have serious disadvantages: a cap is always arbitrary, while the iterative noise explodes quickly:  $\sqrt{\var(x)}\simeq\order(10^3)$ for $\eps=1$ and $i=10$ (and assuming $\Delta=1$ in each case).  Moreover, in both cases, adversaries may try to refresh their $\eps$ budget somehow.

In either case, the situation is more complicated when the noise distribution is not DP-parametrised (\ie $\eps$ and $\delta$ are calculated ad hoc on each single output, rather than being noise input parameters): Then the privacy budget spent must be calculated manually across all outputs to obtain a global $\eps$-DP or $\epsdel$-DP guarantee.

\textbf{Risk-driven vs.\ utility-driven parametrisations}\quad A risk-averse statistics curator may find it desirable to publish only outputs that provide a global $\eps$-DP guarantee.  Manifestly $\eps$-DP output mechanisms may thus be said to follow a {\it risk-driven parametrisation}, because the $\eps$ parameter makes the selected level of risk/utility trade-off transparent.  However, such risk-driven DP parametrisations have their own pitfalls: as noted in~\cite{rinott2018}, the total privacy budget (value of the global $\eps$) is not easily fixed from statistical and/or disclosure control decisions, and the noise scaling is over-protective; both issues will be picked up again in section~\ref{risk_DP}.  Moreover, as shown in section~\ref{pre_noise}, unbounded noise is unavoidable if strict $\eps$-DP is sought.

On the other hand, a {\it utility-driven parametrisation} is much closer to the amount of noise actually injected to each single output count.  For instance, if the noise variance $V$ and its bound $E$ are the input parameters, like in the cell key method, users get a very transparent idea of what happened to the data:  they know the typical noise size $\pm\sqrt{V}$ and that each individual count is at most $\pm E$ off.\footnote{The $E$ parameter is usually not disclosed exactly, because its knowledge gives additional---theoretical---disclosure risks.  However, a vague communication \eg that $E\lesssim 10$ is still very useful (\cf section~\ref{utility_lau}).\label{fn_noise_bound}}
This gives strong utility guarantees on the output, which a risk-driven parametrisation simply cannot provide.  We argue in section~\ref{risk_BN} that risks are well controllable for a static (census-like) output with utility-driven parametrisation.

\section{Risk aspects}\label{risk}

Most direct disclosure risks from population statistics relate to small counts in frequency tables that would reveal unique characteristics of a small group or single persons.  These characteristics could then be used to re-identify individual persons and possibly learn new details about them from the output.  Hence traditional SDC methods tend to focus on such small counts only,\footnote{E.g.\ suppression or rounding of small counts, topcoding or general recoding of rare attributes.\label{fn_sdc_trad}}
but powerful theoretical results have meanwhile shown that entire microdata data\-bases can often be reconstructed accurately from too detailed outputs~\cite{dinur2003}, thus exposing rare records even if small output counts were treated.

Therefore, many current discussions about risk scenarios and subsequently required protection circulate around database reconstruction.  This means heavy computation machinery is used to reconstruct the complete input microdata database from (too) detailed output tables with too little noise.  The procedure is highly effective as described illustratively in~\cite{garfinkel2018}, which assumes that only small counts were treated.  The authors do state, however, that random noise injection is effective, while the database reconstruction theorem~\cite{dinur2003} outlined in section~\ref{pre_db_recon} sets a scaling law for noise magnitude vs.\ output complexity.
Often the theorem and implied reconstruction risks are cited to make a case for manifestly $\eps$-DP outputs~\cite{abowd2018,ashgar2019}.  The next two sections will discuss implications of some classes of output mechanisms introduced in section~\ref{pre_output} on database reconstruction risks, before addressing in section~\ref{risk_re-id} a more fundamental, but no less important question.

\subsection{Risk-driven parametrisation: \texorpdfstring{$\eps$}{eps}-DP mechanisms}\label{risk_DP}


The distinct attractiveness of differential privacy lies in its personal guarantee to each potentially affected individual, which is designed to quantify and limit individual risk at the source and thus forestall any discussions about database reconstruction and its implications.  This formal privacy guarantee holds without making constraints or assumptions on the number or nature of the output mechanism and thus claims to break the \lquote{propose-break-propose cycle} of traditional (\ie non-DP) disclosure control approaches~\cite{dwork2011}.  But how exactly do $\eps$-DP output mechanisms dodge the database reconstruction theorem (section~\ref{pre_db_recon})?

Recall the theorem stating that the input database (size $n$) to any output with bounded noise $E<\order(\sqrt{n})$ and $\poly(n)$ complexity can be reconstructed to high accuracy.  However, it is easy to see that also unbounded (incl.\ $\eps$-DP) noise {\it distributions}, as introduced in section~\ref{pre_noise}, are susceptible in principle: Just pick a small enough width (\ie large enough privacy budget $\eps$ spent on each noise term~$x$) such that all $|x|\leq E_\alpha<\order(\sqrt{n})$ for all the $t\simeq\poly(n)$ required outputs at a set confidence level $\alpha$, and the theorem plays out.  A short calculation using $x\sim\lap(1/\eps)$ puts a lower limit on $\eps$ per noise term as
\begin{equation}\label{eq_eps_a}
    \eps_{\alpha}(n) > \frac{1}{\sqrt{n}}\log\left( \frac{n\log^2 n}{1-\alpha} \right)
\end{equation}
for $t=n\log^2n$ outputs.  For instance, $\eps_{99}>0.024$ per query leads to non-privacy of a database with $n=10^6$ records in $99\percent$ of the cases.
Of course, the global privacy guarantee of such a {\it mechanism} (section~\ref{pre_output}) would be $t\eps$-DP and thus meaningless; the point is that unbounded noise with adjustable variance does not evade the theorem in principle.

Recall the other important result of~\cite{dinur2003}, namely the more general scaling law that noise of size $\order(\sqrt{t})$ suffices to have private outputs of complexity $t$.  So in fact, the key is not the unboundedness of $\eps$-DP {\it noise distributions} but the built-in noise scaling of $\eps$-DP {\it output mechanisms} through the composition theorem~\cite{dwork2014}.  However, note that the $\eps$-DP distribution $\lap(1/\eps)$ (and its discrete sibling from the two-tailed geometric distribution) scales too steeply with $t$:  for a global $\eps$-DP guarantee on an output of complexity $t$, the DP composition rule requires that the global privacy budget be split between all outputs, \eg as $\eps/t$.  Then the noise scales as
\begin{equation}
    \sqrt{\var(x)} = \sqrt{2}\frac{t}{\eps} \sim t
\end{equation}
and not as $\sim\sqrt{t}$, which would be sufficient from the theorem's perspective.  Therefore, $\eps$-DP mechanisms relying on these noise distributions will over-protect outputs of increasing complexity.\footnote{Theorem~3.20 of~\cite{dwork2014} provides for improved composition scaling $\sim\sqrt{t}$, but the resulting global guarantee is relaxed $\epsdel$-DP only, not strictly $\eps$-DP.  Similarly, Gaussian DP noise scales $\sim\sqrt{t}$~\cite{dwork2006a,dwork2016}, but also there the guarantee is only $\epsdel$-DP.\label{fn_advanced_comp}}

There is another issue: the scaling law can be read backwards to note that, whenever the output complexity $t$ is fixed, as \eg in static outputs, also the amount of noise can be fixed globally at the time of output curation.  The order of magnitude should again be $\order(\sqrt{t})$, but irrespective of any assumption on the particular noise distribution, and no DP-specific complexity scaling is needed anymore.  This means in static outputs, like the typical scenarios in official population statistics, any suitable noise distribution can be selected and tuned to size, while DP-specific scaling properties do not add immediate value.  The next sections will shed more light on this.

\subsection{Utility-driven parametrisation: bounded noise}\label{risk_BN}

While from a database reconstruction perspective strictly $\eps$-DP and other, non-DP or $\epsdel$-DP, mechanisms are on level ground at least in static output scenarios, there are strong indications that the latter are more beneficial from a utility perspective, especially if bounded noise distributions are employed as \eg in the cell key (CK) method~\cite{fraser2005,marley2011,thompson2013} (\cf section~\ref{utility}).  However, increased disclosure risks exploiting the tight noise bound were recently claimed for a generic bounded noise mechanism~\cite{ashgar2019}, so a closer look at those results will be insightful.  Ultimately the attacks aim at removing noise from output counts: this boils down to database reconstruction, which is more and more straightforward when less and less noise is present.
In a reasonably prudent scenario, the noise bound $E$ should be non-public, conveying strong utility guarantees solely by publishing the noise variance parameter $V$ together with a vague $E$ limit, \eg $E\lesssim 10$ (a utility-driven parametrisation). Therefore, the first attack in~\cite{ashgar2019} aims at disclosing the exact value of~$E$.

\begin{figure}[t]
    \centering
    \includegraphics[width=0.6\textwidth]{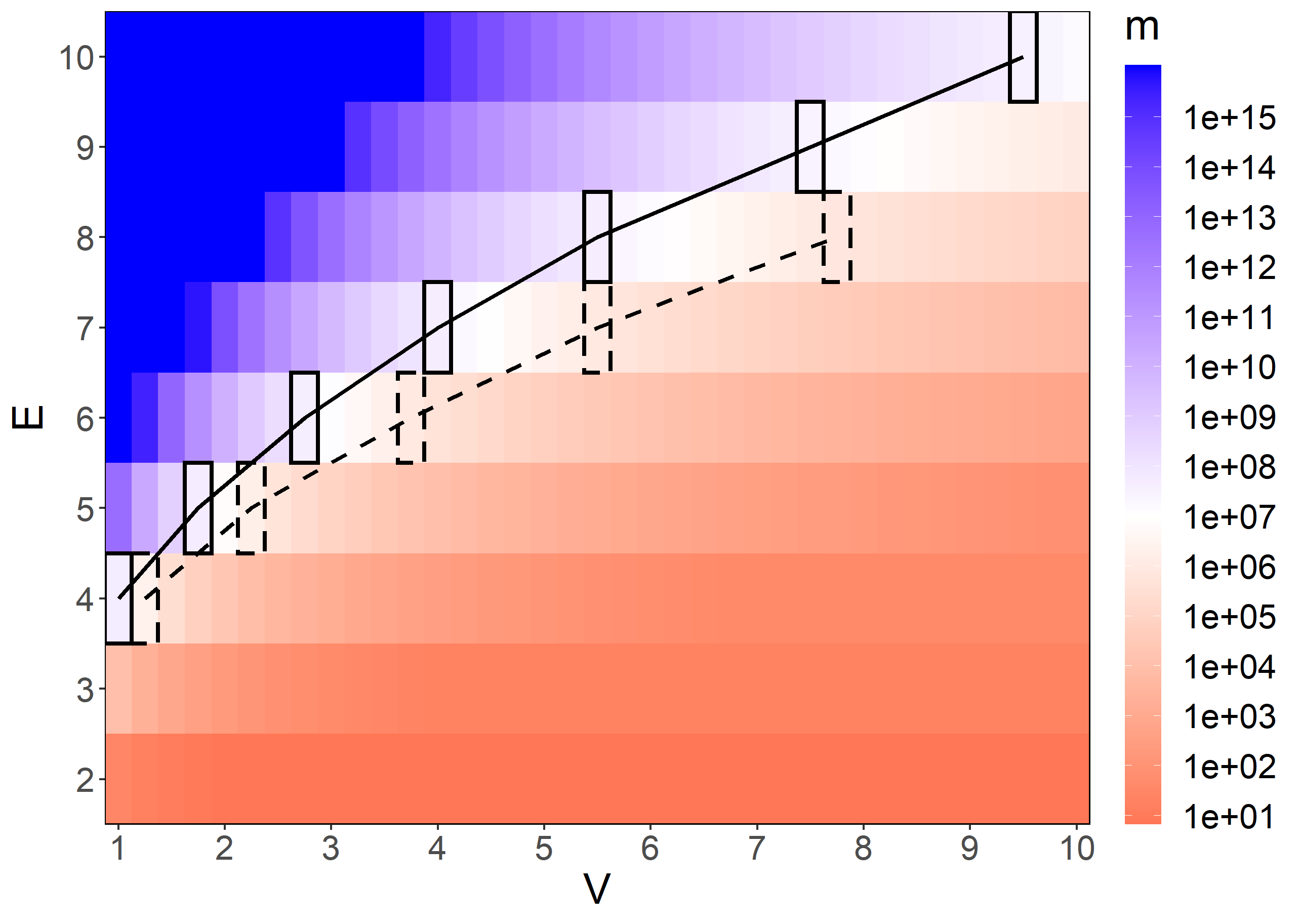}
    \caption{
    Heat map showing the number $m$ of 3-tuples required to infer $E$ at confidence level $\alpha=68\percent$ over the $V$--$E$ parameter space of CK-like methods (and $p_1\simeq 1/m$, \cf Eq.~\eqref{eq_m}).  Black boxes highlight the parameter settings where $m$ exceeds the number of independent 3-tuples (\ie sex breakdowns) available in the 2021 EU census output of Germany (solid) and Malta (dashed).
    }
    \label{fig_risk_BN}
\end{figure}

\textbf{Revealing the bound}\quad The attack in~\cite{ashgar2019} relies on $m$ output 3-tuples of noisy observations with independent noise but respecting a linear constraint, say $\{F,M,T\}$ with expectation $\mathrm{E}(F+M-T)=0$ so that $F+M-T$ values are sampling the noise distribution.\footnote{The type of 3-tuples is not important; we simply choose symbols $F$, $M$ and $T$ to suggest a typical sex breakdown which is almost always available in any population statistics, and to make contact with section~\ref{utility_lau}.\label{fn_3-tuples}}
This gives an estimator for the noise bound
\begin{equation}\label{eq_estimate_E}
    E = \ceil[\big]{\left|\frac{F+M-T}{3}\right|},
\end{equation}
where the probability of revealing $E$ correctly from a single 3-tuple is fixed by the noise distribution as $p_1\define\pr[(F+M-T)>3(E-1)]$.\footnote{E.g.\ uniform noise $\in\{-E,E\}$ gives $p_1=20/(2E+1)^3$ by simple combinatorics~\cite{ashgar2019}.\label{fn_p1_uniform}}
Given~$p_1$, the number of independent 3-tuples needed to infer $E$ at confidence level~$\alpha$ is
\begin{equation}\label{eq_m}
    m = \ceil[\big]{\frac{\log(1-\alpha)}{\log(1-p_1)}} \simeq \ceil[\big]{\frac{1}{p_1}}\quad \text{for}\quad \alpha=68\percent\text{ and } p_1\ll 1.
\end{equation}
Results in~\cite{ashgar2019} are for uniform noise only, but in general $m$ will depend heavily on $p_1$ and thus on the particular noise distribution.  For instance, in CK-like methods $p_1$ is fixed by the \ptable~\cite{thompson2013} and thus by $V$ and $E$ parameters, which allows to control the required complexity~$m$.
Fig.~\ref{fig_risk_BN} illustrates
$m$ over the typical $V$--$E$ parameter space in a generic CK setup using the \ptable~tool recommended for the 2021 EU census~\cite{essnet2019a,meindl2019}.\footnote{
The setup is `generic' because we use the implemented generic \ptable\ generating algorithm that maximises entropy under the sole constraints of fixed $V$ and~$E$ (\cf\cite{giessing2016}), where resulting \ptable s are very close to a Gaussian distribution (see \eg\cite{rinott2018}).  If \ptable s are further tailored to specific needs, \eg adding more constraints, this may affect $p_1$ and thus $m$ as well as resulting quantitative conclusions in this section.\label{fn_CK_EU2021-census}}

Note that $m$ converges to the uniform limit for increasing $V>E$ (because the \ptable\ converges to the uniform distribution with maximum variance $V=E(E+1)/3$), but diverges quickly for decreasing $V<E$ (because large noise magnitudes become increasingly unlikely).  This suggests that CK setups with moderately large $E\lesssim 10$ and considerably smaller $V$ (\eg $E=5$ to 10 and $V=2$) perform as ``quasi-unbounded'' noise on attempts to disclose the exact $E$~parameter.  In conclusion, \cite{ashgar2019} have argued that $E$ cannot be sufficiently protected, but it was shown above that this depends critically on the noise distribution and relative choice of $V$ and $E$: while uniform noise seems $E$-disclosive, generic bounded noise distributions, manifestly $\epsdel$-DP or not, can be set up to protect $E$ effectively while keeping strong utility guarantees (moderate variance and hard noise bound).

\textbf{Exploiting margins}\quad Nevertheless, assume now $E$ is known to complete the discussion.  Then one can search the whole output for constraint n-tuples with extreme noise combinations, which can only be obtained by a single noise pattern applying $\pm E$ to each count.  In such a case, all true counts of the n-tuple are disclosed: \eg find $F=3$, $M=2$ and $T=11$ with $E=2$ known, which discloses true $F=5$, $M=4$ and $T=9$.\footnote{There are more disclosive patterns when true 0s are not perturbed~\cite{enderle2020}, but these require very specific true count patterns combined with a specific noise pattern drawn; such patterns become very unlikely for moderately large~$E$, as suggested in~\cite{enderle2020}.\label{fn_ex_E-tuple}}
However, the abundance of such noise combinations in the output depends again on $p_1$ described above (or its generalisation for $>2$ categories), and thus becomes increasingly unlikely in the ``quasi-unbounded'' regime ($V\ll E\lesssim 10$): The authors of~\cite{bailie2019} estimate such risks in a typical scenario as $\order(10^{-3}-10^{-16})$, but assuming $E=2$ fixed, while similar risk scaling with $E=1$, 2 or~5 (for fixed~$V$) is observed in~\cite{enderle2018}.  Fig.~\ref{fig_risk_BN} also shows $p_1\simeq 1/m$ for a two-category variable (most disclosive) as a function of $E$ and~$V$.  For instance, with $E=5$ and $V=2$ each output 3-tuple has just an individual chance $p_1\simeq\order(10^{-7})$ of being divulged, and for $E=10$ and $V\leq 4$ it is practically zero.  In section~\ref{risk_av} we count the number of independent sex breakdowns available in the 2021 EU census output: this depends on the geographic breakdowns and hence on country size, giving the largest available $m=2.8\times 10^7$ for Germany.\footnote{The counting in section~\ref{risk_av} does not include the $1\,\mathrm{km}^2$ grid output (\cf annex~\ref{a_eu_census_sdc}), but the additional number of sex breakdowns (one per grid cell) is negligible here: \eg for Germany the grid would provide another $3.6\times 10^5$ 3-tuples, a $1\percent$ effect.\label{fn_3-tuples_grid}}
Note finally that such an attack cannot be ``aimed'' at specific observations of interest; it is limited to wherever extreme noise patterns happen to occur.  Targeted attacks must pursue other strategies, addressed in section~\ref{risk_av}.

\textbf{Heuristic parameter constraints}\quad To generalise these results, a heuristic risk constraint can be inferred on the $V$--$E$ parameter space of static outputs: to avoid $E$-disclosure, choose $V$ and $E$ such that the $E$ disclosure risk is below $68\percent$, even when all available 3-tuples are used.  The respective contours are added to Fig.~\ref{fig_risk_BN} for Germany (most independent sex 3-tuples) and for Malta (fewest independent sex 3-tuples).  In general, such a limit can always be set from the number of independent 3-tuples in the static output and $p_1$ of the (bounded) noise distribution.  Note that this constraint is very conservative: even if $E$ was disclosed correctly, the number of 3-tuples with noise possibly removed would remain in the single digits.
The next section will put an independent limit on $V$ for static outputs, irrespective of the specific noise distribution or output mechanism.

\subsection{All noisy mechanisms: Massive averaging}\label{risk_av}

The second step of the attack in~\cite{ashgar2019} aims to remove the noise successively from certain (sets of) table cells (called `histogram reconstruction').  The concept underlying such attack classes is {\it massive averaging}, to which any noise method with constant variance is susceptible.  This is illustrated by Chebyshev's inequality (its inverse, to be precise):
\begin{equation}\label{eq_cheb}
    \pr(|\overline{x}|<\xi) \geq 1-\frac{\kappa\var(x)}{\xi^2 t},
\end{equation}
where $\overline{x}$ is the averaged (unbiased) noise over $t$ independent observations of the same statistic of interest, $\var(x)$ the variance of each individual noise term, $\kappa$ a constant factor counting how many outputs had to be summed on average to obtain the statistic (\eg $\kappa=2$ for $t$ bi-partitioned observations), and $\xi$ a small parameter (\eg $\xi=0.5$ for accurate disclosure of an integer count~\cite{ashgar2019}).  The $t$ scaling comes from the averaging, so if $\var(x)$ is constant in $t$, the average always becomes accurate for sufficient~$t$.  More precisely, Eq.~\eqref{eq_cheb} implies that the noise must scale as $\var(x)\sim t$, \ie noise of~$\order(\sqrt{t})$, to prevent this.  This is consistent with the scaling law in~\cite{dinur2003} (\cf \cite{ashgar2019}).

\textbf{Counting redundancies}\quad However, massive averaging requires enough redundant representations of the same target statistic in the output, but with independent noise such that $\overline{x}$ is random and Eq.~\eqref{eq_cheb} plays out.  For instance, \cite{ashgar2019} rely on $t$ {\it user-defined} independent bi-partitions of the same variable.  Then the noise of any target statistic broken down $t$ times by these bi-partitions can be removed by averaging over $t$ sums of the bi-partitions.  Note that only the SPSN principle (section~\ref{pre_output}) assumed here to be present requires the use of n-partitions in the first place; without SPSN the adversary could just query $t$ times the target statistic directly.  Clearly the scenario in~\cite{ashgar2019} is an example of a badly curated flexible output mechanism.  The situation is quite different with a {\it static} output mechanism as defined in section~\ref{pre_output}: in this case the independent redundant representations (IRR) of each output statistic can be counted in advance, and curated if needed.  The output complexity can still be sizeable, and table builders giving users some flexibility can be the preferred publication mechanism.
\begin{table}[t]
\centering
\begin{tabular}{ c||c|c|c||c|c|c| }
    & \multicolumn{3}{c||}{With SPSN} & \multicolumn{3}{c|}{Without SPSN} \\
    $T_\mathbf{A}$ & $t$ & $k$ (DE) & $k$ (MT) & $t$ & $k$ (DE) & $k$ (MT) \\
    \hline
    total            & 2775 & $8.6\times 10^7$ & $2.1\times 10^6$ & 6378 & $9.7\times 10^7$ & $2.6\times 10^6$ \\
    SEX              & 1376 & $2.8\times 10^7$ & $7.1\times 10^5$ & 3105 & $3.2\times 10^7$ & $8.4\times 10^5$ \\
    AGE.M            &  926 & $3.0\times 10^6$ & $7.9\times 10^4$ & 2281 & $3.4\times 10^6$ & $9.7\times 10^4$ \\
    GEO.L            &  766 & $5.4\times 10^5$ & $5.4\times 10^5$ & 1734 & $6.6\times 10^5$ & $6.6\times 10^5$ \\
    SEX\,AGE.M &  458 & $9.6\times 10^5$ & $2.6\times 10^4$ & 1097 & $1.1\times 10^6$ & $3.2\times 10^4$ \\
    \hline
\end{tabular}
\caption{The top five output statistics $T_\mathbf{A}$ with the largest number $t$ of IRRs in the 2021 EU census output; $k$ depends on the geographic breakdowns, so we list largest (Germany) and smallest (Malta) results.}
\label{tab_risk_BN_2}
\end{table}

\textbf{2021 EU census example}\quad To understand what controlling redundancies in a static output means, it is insightful to do the full exercise on the 2021 EU census setup.  Annex~\ref{a_av} describes step by step how all IRRs of every statistic contained in an output of many contingency tables can be counted systematically, with and without SPSN invoked.  Eq.~\eqref{eq_cheb} suggests $\kappa/t\equiv k/t^2$ as an averaging risk measure, where $t$ is the number of IRRs being averaged and $k$ is the total number of independent counts (\ie noise terms) contributing to the IRR average.  Both $t$ and $k$ are obtained from the counting exercise, for every statistic contained in the output.\footnote{{\it Example:} Consider a single two-way output table $\text{SEX}\times\text{AGE}$ with $\text{SEX}=\{F,M,T\}$ and $\text{AGE}=\{<30,\geq 30,T\}$, \ie including all margins ($T$) and thus 9 independent cells.  Here the total population count comes in $t=4$ IRRs: the trivial total margin ($k=1$), two sums over the SEX/AGE margins ($k=2$ each), and a sum over all internal counts ($k=4$), so total $k/t^2=9/16$.  The same analysis is straightforward for all other statistics covered: counts broken down by AGE {\it or} SEX ($t=2$ and total $k=3$ each, so $k/t^2=3/4$) as well as counts by AGE {\it and} SEX ($t=1$ and $k=1$ each, so $k/t^2=1$).  In annex~\ref{a_av}, this counting analysis is done systematically for all statistics contained in the full 2021 EU census output.\label{fn_IRR_example}}
Table~\ref{tab_risk_BN_2} lists the top five output statistics by number $t$ of IRRs available, and Table~\ref{tab_risk_BN} (in annex~\ref{a_av_census}) shows the top-five smallest $k/t^2$ values found, together with the respective output statistic and with/without SPSN.
Fig~\ref{fig_risk_BN_2} (in annex~\ref{a_av_census}) shows the distribution of $k/t^2$ across all output statistics without geographic dependence in their IRRs.  While a comprehensive discussion of the results is provided in annex~\ref{a_av_census}, it suffices to note here that SPSN is an effective means of reducing $t$ and thus lowering averaging risks, \cf Eq.~\eqref{eq_cheb}.

\begin{figure}[t]
    \centering
    \includegraphics[width=0.49\textwidth]{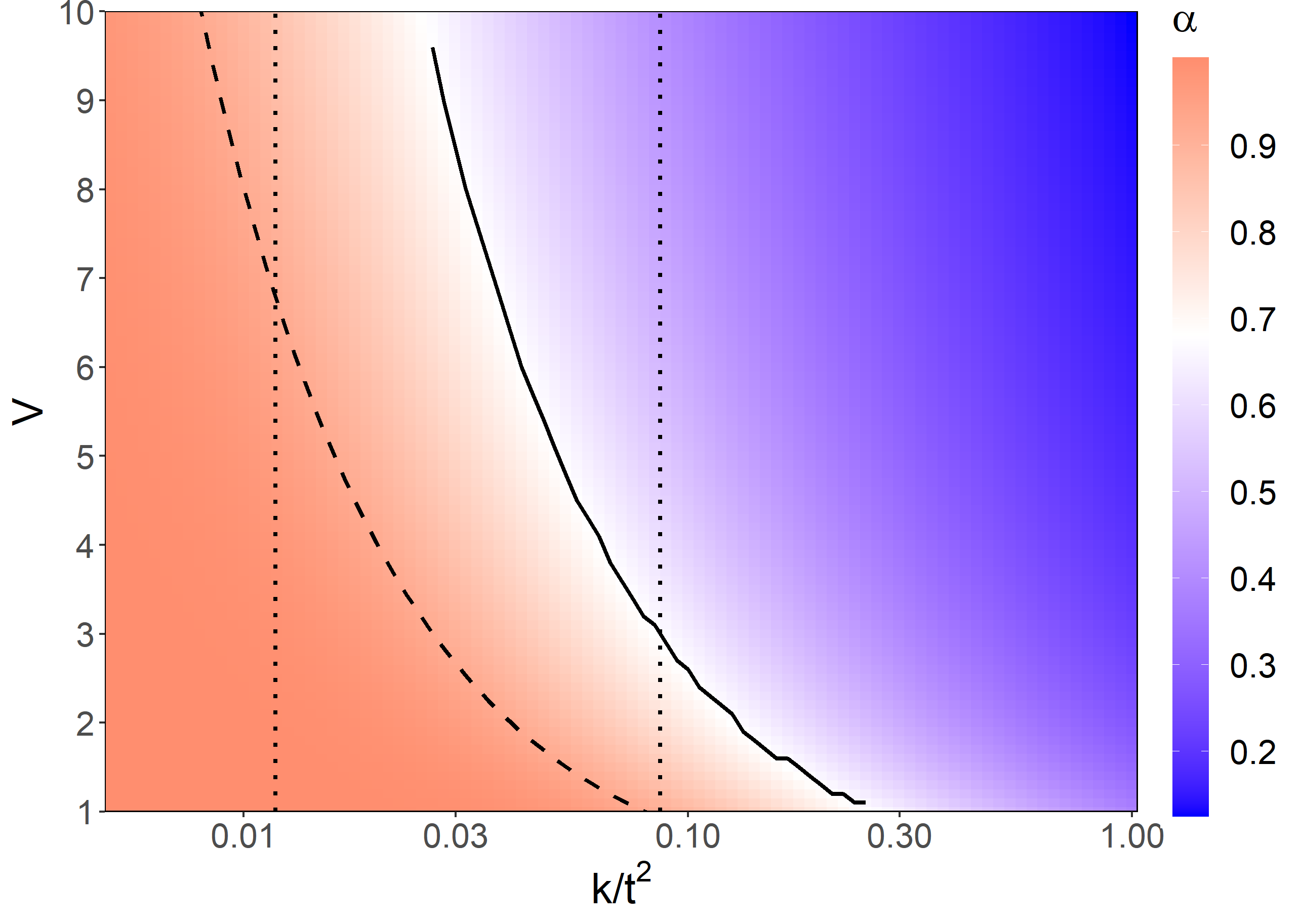}
    \includegraphics[width=0.49\textwidth]{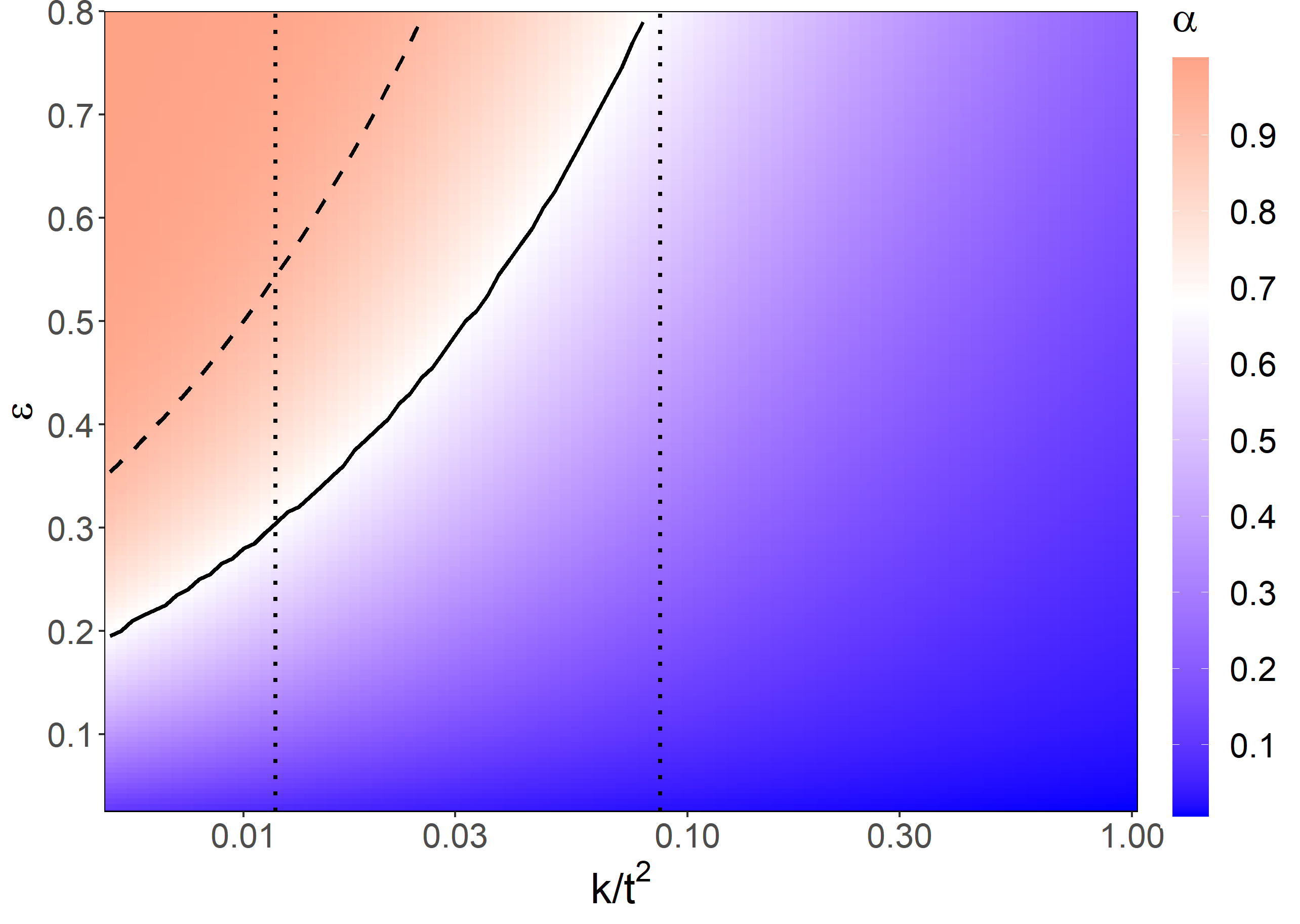}
    \caption{Heat maps of the $k/t^2$ vs.\ noise parameter space showing the Gaussian-modelled probability $\alpha$ of averaging a single output count correctly, for the CK variance parameter $V$ (left) and the DP privacy budget $\eps$ per single count (right).\newline
    Both plots also show the Gaussian $\alpha=68\percent$ contour (bold line) and Chebyshev's lower limit at $\alpha=68\percent$ (dashed line), as well as the smallest optimised $k/t^2$ values found in the 2021 EU census output with and without SPSN (dotted lines; \cf Table~\ref{tab_risk_BN} in annex~\ref{a_av_census}).}
    \label{fig_risk_av}
\end{figure}

\textbf{Heuristic variance constraint}\quad One could confront the smallest $k/t^2$ values in the output with Eq.~\eqref{eq_cheb} for $\xi= 0.5$ to obtain a limit on $V$. However, Eq.~\eqref{eq_cheb} represents a lower limit on the averaging success, so for a more conservative constraint we assume Gaussian-distributed noise sums, Eq.~\eqref{eq_gaussian_var}; see annex~\ref{a_av_risk} for a detailed reasoning.
Fig~\ref{fig_risk_av} shows that a CK setup with $V\simeq 3$ and SPSN is sufficient to reduce the risk of averaging correctly even the most risky output statistic to below $68\percent$ per count.  Note again that this is a very conservative constraint:  the chances of obtaining correct averages for $m$ output counts would shrink as $\alpha^m$, and $k/t^2$ is generally larger.\footnote{According to Table~\ref{tab_risk_BN} in annex~\ref{a_av_census}, the AGE.M histogram for Luxembourg is with $k/t^2=0.0867$ the most risky statistic in the whole 2021 EU census output.  AGE.M has 21 categories, so the chance of averaging the entire histogram correctly would be $\leq 0.68^{21}\lesssim 0.03\percent$.\label{fn_av_LU}}
Again the approach can be generalised to any static output: just measure the most risky output statistic with $k/t^2$ and fix the variance from Fig.~\ref{fig_risk_av} (or reduce output complexity).

\textbf{The $\eps$-DP picture}\quad Note first that the global privacy guarantee would be $t\eps$-DP, with $t$ the number of IRRs of the population total (\cf Table.~\ref{tab_risk_BN_2}) and $\eps$ the budget spent on each IRR.  This is easy to see: the whole output can be generated by calling the output mechanism once asking for all IRRs of the `total' statistic.  The global sensitivity, Eq.~\eqref{eq_df}, of this query is $\Delta=t$, with $t$ the number of IRRs of the population total, because adding or removing a single person would change exactly one count in every IRR by~1.  Hence each IRR should obtain count-level noise $x\sim\lap(t/\eps_\mathrm{global})$ to establish an $\eps_\mathrm{global}$-DP guarantee (\cf section~\ref{pre_output}).  Seeing $t=2\,775$ with SPSN (and $t=6\,378$ without), this is obviously an extreme scenario leading to count-level noise of size $\sqrt{\var(x)}\simeq\order(10^3-10^4)$ for $\eps_\mathrm{global}\simeq\order(1-10)$.  However, section~\ref{risk_DP} suggests that the composition rule underlying strict $\eps$-DP mechanisms overprotects large-complexity outputs.  On the other hand, an averaging constraint can put a bottom-up limit on count-level $\eps$ budgets: the Gaussian model\footnote{Clearly a sum of $k$ Laplacian noise terms will not be Gaussian-distributed; however, the approximation will not be orders of magnitude off, and still better than the Chebyshev lower limit.  In fact, a similar test as for the CK \ptable\ noise described in footnote~\ref{fn_ck_gauss_test} yields good agreement between $10^3$ $\lap(1/\eps)$ samples and the Gaussian estimate with $V=2/\eps^2$.}
in Fig~\ref{fig_risk_av} suggests $\eps\lesssim 0.3$ without SPSN (the standard DP lore; with SPSN it would be $\eps\lesssim 0.8$).

\subsection{Database reconstruction vs.\ personal data breaches}\label{risk_re-id}

After looking into various database reconstruction risks and preemptive measures to control them, the section concludes by addressing an important, more fundamental question~\cite{ruggles2019}:
\begin{displayquote}
What does database reconstruction actually mean for disclosing personal information of any natural person?
\end{displayquote}
The answer in~\cite{ruggles2019} is: A priori not much.  It is argued that one still has to establish a {\it reliable connection}, using auxiliary external information, between a given database record and an actual person.  On the other hand, it is information on a {\it natural person} (`personal data' of a `data subject') that is protected by modern data protection standards like the GDPR\footnote{Regulation (EU) 2016/679 of the European Parliament and of the Council of 27 April 2016 on the protection of natural persons with regard to the processing of personal data and on the free movement of such data [\ldots] (General Data Protection Regulation) (\href{https://eur-lex.europa.eu/legal-content/EN/TXT/?qid=1578673009767&uri=CELEX:32016R0679}{OJ~L~119, 4.5.2016, p.~1}).\label{fn_gdpr}} (\cf definitions in Article~4).
From this perspective, database reconstruction may be seen as the first step of a fully fledged re-identification attack aiming to establish a {\it correct one-to-one link} between information contained in a given statistical output and a natural person, or data subject (an instance of a `personal data breach' in GDPR terms).\footnote{In fact, the related discussion in the U.S.\ on 2020 census protection has a legal dimension addressing what exactly is protected by U.S.\ privacy law~\cite{ruggles2019,mervis2019}, which may have been fuelled also by a political controversy on the questions to be included in the 2020 U.S. census~\cite{mervis2018}.  In any case, all statements here on EU privacy law purely reflect the authors' personal perceptions and do not forestall any formal or authoritative legal interpretation of applicable EU law.\label{fn_legal_disclaimer}}

In practice, re-identification attempts will be probabilistic exercises:
\begin{itemize}
    \item The database reconstruction will have a specific success rate $r_\textrm{recon}$, defined as the share of individual records reconstructed {\it correctly} (\ie matching a record in the input database).  This rate can be controlled by restricting the amount of output statistics published, in combination with post-tabular SDC methods injecting noise in the output mechanism (section~\ref{pre_output}).
    \item Then the matching of a correctly reconstructed database record to a natural person will have an independent success rate $r_\textrm{match}<100\percent$, even for unique records reconstructed from census outputs with full enumeration.  This is due to intrinsic inaccuracies in the input database and in the auxiliary external information used for the matching, but it can be controlled further by pre-tabular SDC methods adding noise to the input microdata (\eg record swapping).
\end{itemize}
In consequence, the re-identification success rate---quantifying the likelihood of a successful re-identification of an actual natural person, \ie personal data breach in the GDPR sense---is
\begin{equation}\label{eq_r}
    r_\textrm{re-id} = r_\textrm{recon} \cdot r_\textrm{match},
\end{equation}
where both independent rates on the right-hand side are intrinsically $<100\percent$\footnote{For instance, the \uscb\ has staged a re-identification attack on its on published 2010 U.S.\ census data, finding $r_\textrm{recon}\sim 50\percent$ but only when also including hits with some record variables (\eg age) a bit off~\cite{mervis2019,hansen2018}.  The subsequently attempted matching to real persons seems to have returned a sufficiently low  $r_\textrm{match}$ to lead the authors themselves to the conclusion that \lquote{the risk of reidentification is small} (\cite{abowd2018} p.~15).\label{fn_us2010_db_recon}} 
and can be further controlled by SDC methods, as outlined above.

In any case, the fundamental SDC law holds that no output utility is for free.  As argued in section~\ref{risk_DP}, differential privacy steps in line (recall: ``no outputs [\ldots] would become {\it significantly} more or less likely''), but adds risk transparency with a privacy budget parametrisation, \ie tuning what ``significantly'' means.  Therefore, irrespective of a particular mechanism, any useful statistical output will have a trailing re-identification risk $0<r_\textrm{re-id}<100\percent$, for which an upper bound can be controlled with SDC methodology.

Of course, this does not touch upon the notorious philosophical question what is still an acceptable (upper limit on) $r_\textrm{re-id}$: Is a chance for some rare individuals of 1 in 10, or 1 in 5, or 1 in 4 still considered acceptable?  The GDPR does not answer this from a legal perspective (nor any other relevant data protection legislation to the authors' knowledge), it rather acknowledges such trailing risks and defines ensuing responsibilities of the data controller in paragraph~1 of Article~25 (`Data protection by design and by default'):
\begin{displayquote}
\lquote{Taking into account the state of the art, the cost of implementation and the nature, scope, context and purposes of processing as well as the risks of varying likelihood and severity for rights and freedoms of natural persons posed by the processing, the controller shall [\ldots] implement data-protection principles [\ldots] in an effective manner [\ldots].}
\end{displayquote}
For the purpose of official statistics in the EU, these principles are further specified in Regulation (EC) No 223/2009 on European statistics\footnote{Regulation (EC) No 223/2009 of the European Parliament and of the Council of 11 March 2009 on European statistics (\href{https://eur-lex.europa.eu/legal-content/EN/TXT/?uri=CELEX:02009R0223-20150608}{OJ~L~87, 31.3.2009, p.~164}).\label{fn_reg_223-2009}}, which the GDPR acknowledges in Recital~163 as the relevant legal framework.

It thus remains a key mandate of statistical authorities to assess, quantify and control disclosure risks, and ultimately to make a choice on acceptable risks against remaining usefulness of their products.\footnote{Note the definition of `confidential data' in Art.~3(7) of Reg.\ (EC) No 223/2009, which requires for the assessment of re-identification risk that ``account shall be taken of all relevant means that might {\it reasonably} be used by a third party to identify the statistical unit''.  The formulation is quite similar to its counterpart in U.S.\ legislation (see~\cite{ruggles2019} section~I), giving statistical authorities the necessary leeway to proceed pragmatically based on best available knowledge, or state of the art as termed in GDPR Art.~25(1).\label{fn_confidential_data}}
Differential privacy does not change this game but essentially adds a risk measure and some more SDC mechanisms to the market.  These should be assessed, together with others, for each specific statistics scenario to find the most beneficial setup.

\section{Utility aspects}\label{utility}

Section~\ref{risk} argued that from a risk perspective, strictly $\eps$-DP output mechanisms---requiring unbounded noise distributions---do not offer immediate benefits over mechanisms with utility-driven parametrisations---including bounded noise---for static outputs.  Shifting now to a utility perspective, there are already various studies assessing utility aspects of DP output mechanisms or testing them in statistical applications, see \eg~\cite{machanavajjhala2008,ghosh2012,petti2019,dwork2010a,wang2015}.

A key reference on utility within the scope of this paper is~\cite{rinott2018}, which has the same statistical setting (census-like unweighted frequency tables of higher dimensionality) and provides a comprehensive study of DP mechanisms and utility measures before this background.  The paper analyses effects of DP noise setups from Laplace and Gaussian distributions on various generic risk and utility indicators, measured on a 7-dimensional output table from the 2001 census in the United Kingdom.  Moreover, the paper assesses potential DP noise effects on data correlations by generating dummy tables with Poisson-distributed counts independent from table dimensions, and then testing the independence hypothesis with noise added.  In line with earlier results~\cite{wang2015}, the test power in terms of significance levels is found to be much worse with a na\"ive model (treating the counts as if they were unperturbed) than with a noise-aware model.

However, all DP noise distributions tested in the paper were truncated to obtain bounded noise, so results do not cover any tail effects from unbounded noise.  On the other hand, the \uscb\ foresees unbounded $\eps$-DP noise for its 2020 census~\cite{abowd2018}, which has triggered severe utility concerns~\cite{ruggles2019,santos2020}.  Some utility implications of published test setups were assessed in~\cite{petti2019}, but the authors explicitly mention the issue of possible tail effects that should be assessed.  The further course of this section concentrates on such tail effects.

\subsection{Parameter setups}\label{utility_pars}

In~\cite{rinott2018}, $\epsdel$-DP noise (as introduced in section~\ref{pre_noise}) was modelled using truncated Laplace and Gaussian distributions with $\eps\in\{0.5, 1.5\}$ and $\delta\in\{2\times 10^{-5}, 8\times 10^{-3}\}$, leading to noise bounds $E=7$ to~12. It is stressed there, however, that these setups do not claim to represent any realistic risk/utility compromise; they were simply chosen to set a common benchmark for comparing the different noise models.  Nevertheless, within the scope of bounded noise effects on census-like table outputs, conclusions from~\cite{rinott2018} should hold also for cell key (CK) setups with similar parameters $V\simeq\order(1)$ and $E\lesssim\order(10)$.  Conversely, the tight bound of $\epsdel$-DP noise in~\cite{rinott2018} is as effective against tail effects as CK noise.

Now censuses are among the most expensive national statistical exercises, serving a variety of specific research and policy purposes, so ensuring that SDC methods maintain unique census features is critical.  Therefore, in contrast to the approach of~\cite{rinott2018} (truncating DP noise and fixing reference $\eps$ and $\delta$ values for a theoretical comparison), this paper aims to assess actual parameter setups currently being discussed in different census contexts, and in particular to focus on tail effects of unbounded $\eps$-DP noise on one of the key unique census features: accurate demographic statistics at very high geographic detail.  Before looking into this, the parameter setups for the different noise distributions discussed are briefly introduced.

\textbf{Noise parametrisations}\quad It is important to note that, from a utility perspective, the specifics of noise added at the {\it individual count level} are the key parameters for any kind of serious research~\cite{rinott2018}.  While utility-driven parametrisation provide this out of the box ($V$ values and $E$ ranges), an effort has to be made in DP setups to infer the $\eps$ spent at the individual count level: as outlined in section~\ref{pre_output}, a global $\eps$ must be split across different partially redundant outputs, and the methods for doing this in an optimised manner can become complex and rather non-transparent~\cite{garfinkel2019}.  Nevertheless, we will attempt to make an educated guess at the $\eps$ budget spent on a single output table, which sets the noise variance for each count in that table.

\textbf{2020 U.S.\ census setup}\quad The \uscb\ plans to apply discrete $\eps$-DP noise from the two-tailed geometric distribution~\cite{petti2019}, \cf Eq.~\eqref{eq_geom2}, with a {\it global} privacy budget in the range $\eps_\mathrm{global} \in \{0.25,0.5,1.0,2.0,4.0,8.0\}$~\cite{garfinkel2019,petti2019}.
This global budget is then distributed across six hierarchical geographies~\cite{garfinkel2019}.  Certain optimisations may shift the relative shares away from an even split, but we assume $1/6$ for practical purposes as in~\cite{petti2019}.  Further intricacies include that noisy total population counts are generated for each geographic level\footnote{Except at State level, where the U.S.\ Constitution requires the \uscb\ to publish unperturbed totals~\cite{petti2019}.\label{fn_us_state}}
and all further breakdowns are optimised to sum to those totals~\cite{petti2019}.  The reference also suggests that at each geographic level, $67.5\percent$ of the budget are spent on the more important person aggregate tables.  In summary, we assume
\begin{equation}\label{eq_us_setup}
    \eps_\mathrm{table} = 67.5\percent \times 1/6 \times\eps_\mathrm{global} \simeq 10\percent\times\eps_\mathrm{global},
\end{equation}
so $\eps_\mathrm{table}\in\{0.025,0.05,0.1,0.2,0.4,0.8\}$ for tabular (count-level) $\eps$-DP noise in section~\ref{utility_lau}.  In accordance with $\lap(\Delta/\eps)$ variance and tabular $\Delta=1$, Eq.~\eqref{eq_lap}, this corresponds to noise sizes at single count level of
\begin{equation*}
    V\in\{3200,800,200,50,12.5,3.125\},\quad \sqrt{V}\in\{56.6,28.3,14.1,7.1,3.5,1.8\}.
\end{equation*}

\textbf{2021 EU census setup}\quad In general, the situation for the 2021 EU census is much more diverse, as SDC treatment is within national competence and countries will likely select a variety of different SDC mechanisms and setups.  Nevertheless, The European Statistical System\footref{fn_ess} has developed recommendations for a harmonised SDC approach based on the CK method~\cite{essnet2017,bach2018}.  These recommendations include some indications on the size of $V\simeq\order(1)$ and $E\simeq\order(1-10)$, where results of sections~\ref{risk_BN} and~\ref{risk_av} may give further guidance.  The resulting count-level noise sizes are orders of magnitude smaller than in the strict $\eps$-DP mechanism above (except for the limiting $\eps_\mathrm{table}=0.8$), and no tails effects $>E$ will be present by definition.  Therefore, this setup will not be discussed in the next section.

\subsection{Demographics at high geographic detail}\label{utility_lau}

Accurate demographics at a high geographic detail is one of the key unique census features in many world regions.  For instance, censuses are until now the only source of accurate population statistics at the Local Administrative Unit (LAU) level across Europe.\footnote{The 2021 EU census round will, for the first time, produce key census indicators on a pan-European $1\,\mathrm{km}^2$ grid~\cite{bach2018}; see annex~\ref{a_eu_census_sdc} for a dedicated comment.\label{fn_grid}}
The 2021 EU census round will cover ca.~$110\,000$~LAUs with a total population of roughly $4.5\times 10^8$ people across the whole EU.\footnote{The data underlying this section are 2011 census results from all EU Member States, except Slovenia for which 2011 census data were not available at the time of writing due to a technical problem.}
Coincidentally this matches well with U.S.\ census outputs at tract level, covering ca.~$75\,000$ geographic units~\cite{garfinkel2019} with a total population of $3.3\times 10^8$ people.  However, the following analysis is intended solely to discuss effects of a practical $\eps$-DP noise scenario on key EU census outputs.  Whether any of the conclusions may apply to tract-level U.S.\ census outputs depends critically on the correctness of parameter assumptions, Eq.~\eqref{eq_us_setup}, and also on the comparability of population distributions across EU LAUs vs.\ U.S.\ tracts.

\begin{figure}[t]
    \centering
    \includegraphics[width=0.49\textwidth]{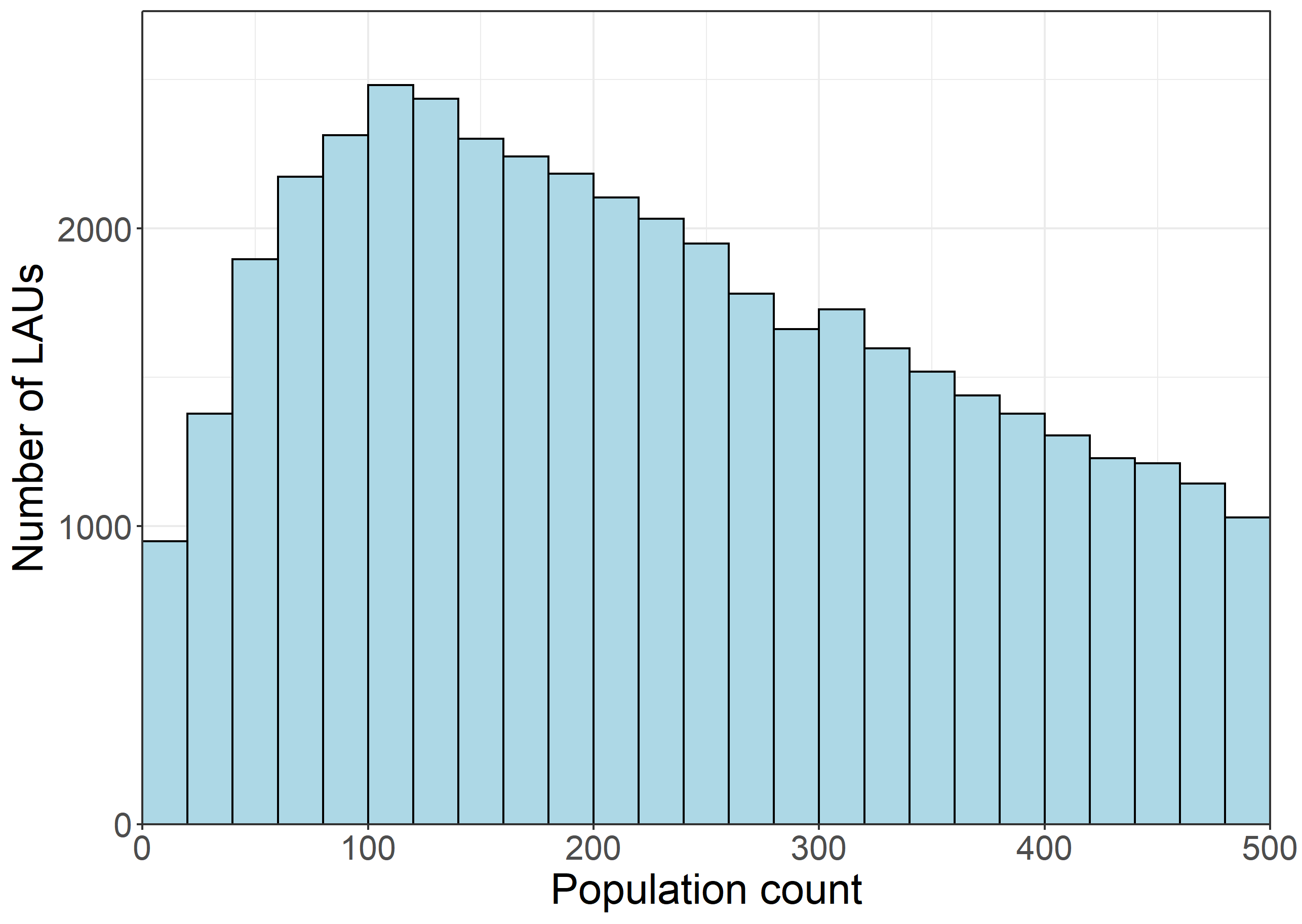}
    \includegraphics[width=0.49\textwidth]{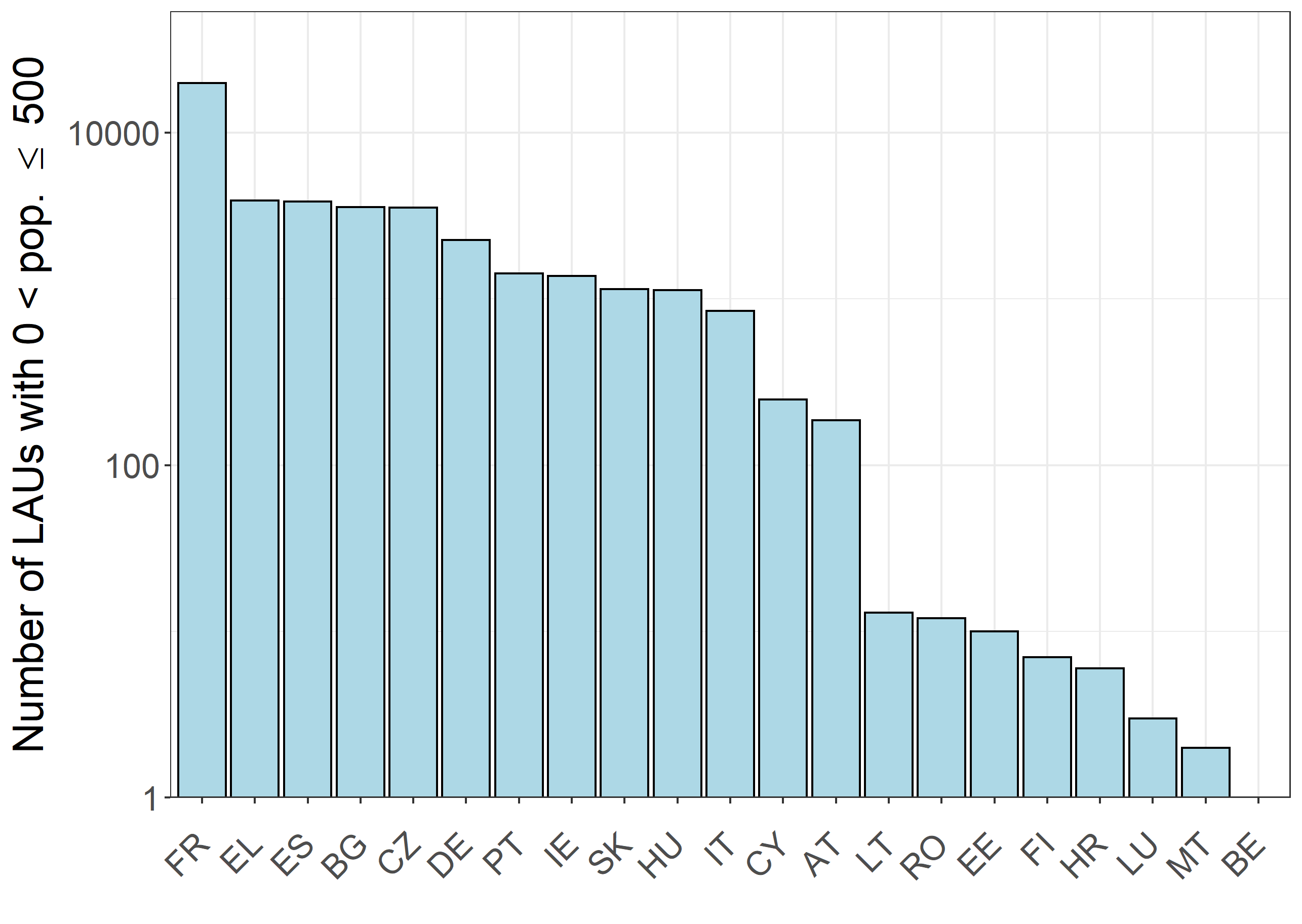}
    \caption{Distribution of populated LAUs with $<500$ residents across the total population count (left) and across EU Member States (right).}
    \label{fig_utility_lau_1}
\end{figure}

\textbf{The statistics of LAUs}\quad There is an extreme variety of total population by LAU, with populated units ranging from $\order(1)$ residents (450 LAUs with $<10$ people) to $3.3\times 10^6$ residents (Berlin; in total 14~LAUs with $>10^6$ people).  Now the key point is that statistics {\it across} LAUs is only part of the purpose of these census results; they are also the only source to obtain accurate demographic information on {\it individual} LAUs.  For this purpose, even very unlikely but very large noise outliers can have severe, maybe unacceptable, consequences.  Furthermore, if the method of adjusting inner tables to their geographic totals after drawing noise is applied~\cite{petti2019}, a single large noise outlier on a given small LAU total would systematically and heavily distort all statistics published for that LAU.  Therefore, the subsequent focus is on LAUs with counts $<500$ illustrated in Fig.~\ref{fig_utility_lau_1}.

\textbf{The demographics of LAUs}\quad To add a demographic element, we include a breakdown by $\mathrm{sex}=\{F,M,T\}$.  This is the spine of all LAU-level person tables in table groups 3 and 8 of the 2021 EU census programme\footref{fn_cir-2}.  It also reflects a possible notion of picking more important `aggregate tables' to which all further breakdowns would then be adjusted~\cite{petti2019}.  To cover both large distortions of totals as well as of sex balances, the counts of $F$, $M$ and $T$ are treated independently.  In total, there are $\sim 167\,000$ observations of $F$, $M$ or $T<500$ at LAU level in the data.

\begin{figure}[!htbp]  
    \centering
    \includegraphics[width=0.49\textwidth]{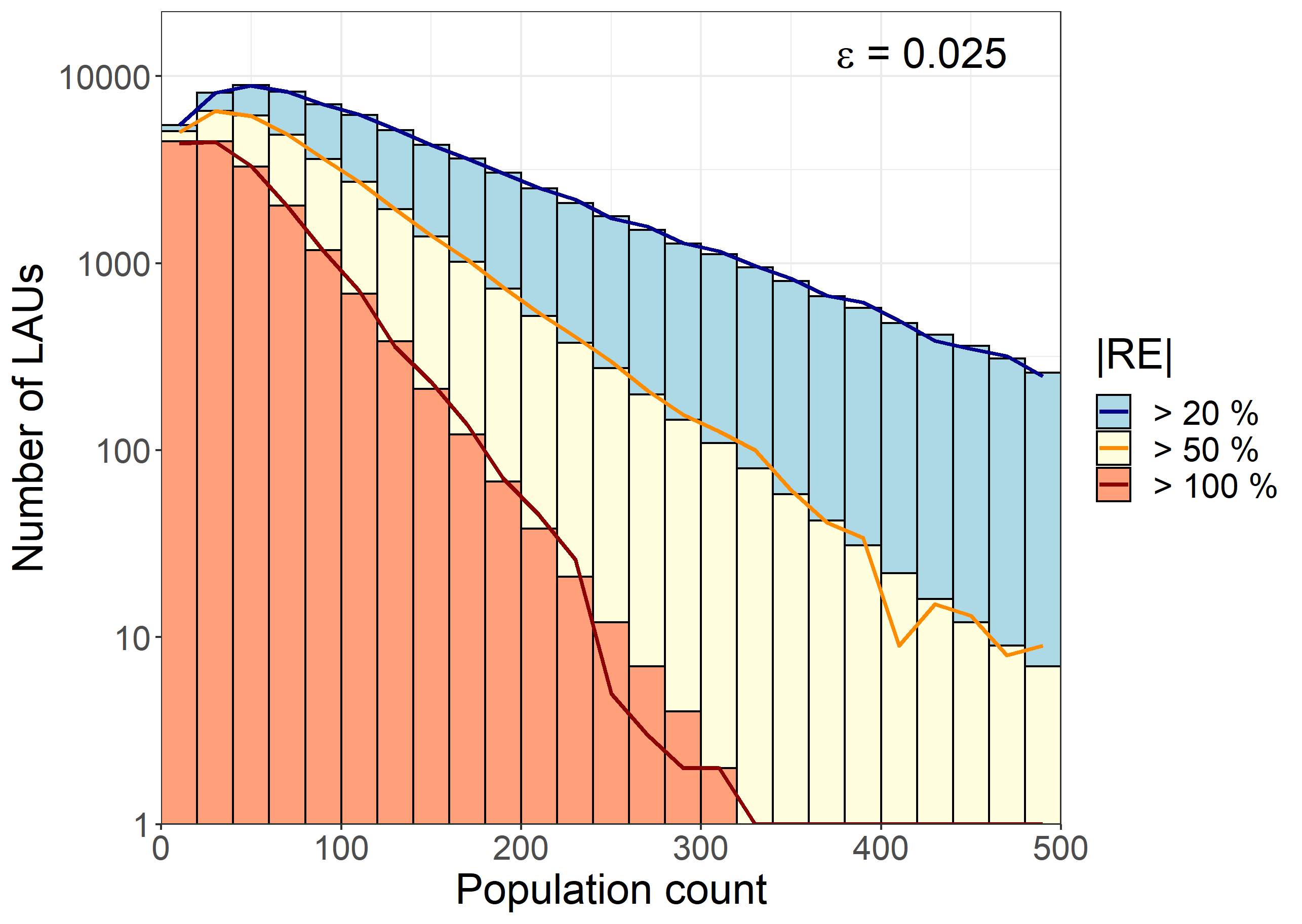}
    \includegraphics[width=0.49\textwidth]{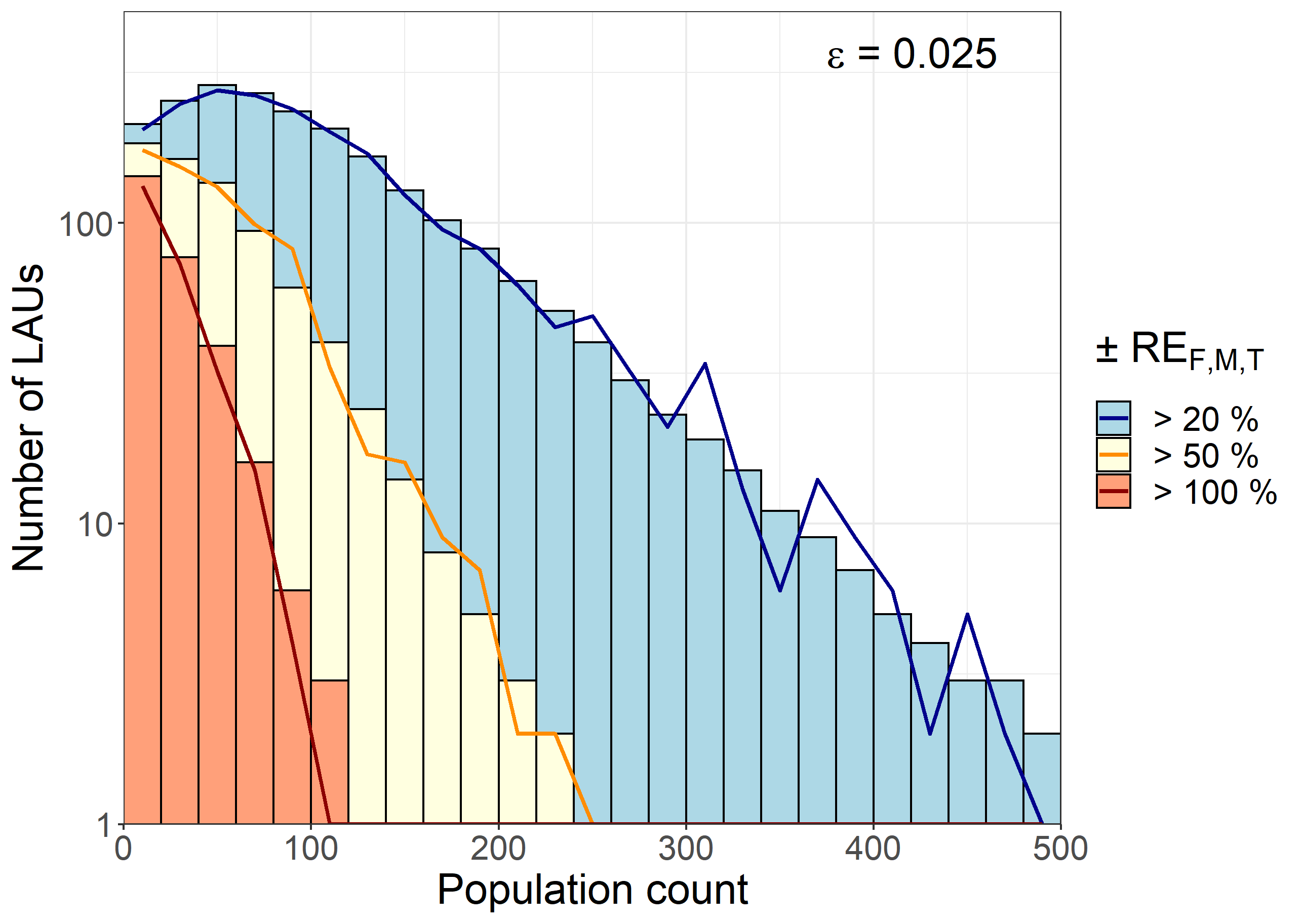}
    \includegraphics[width=0.49\textwidth]{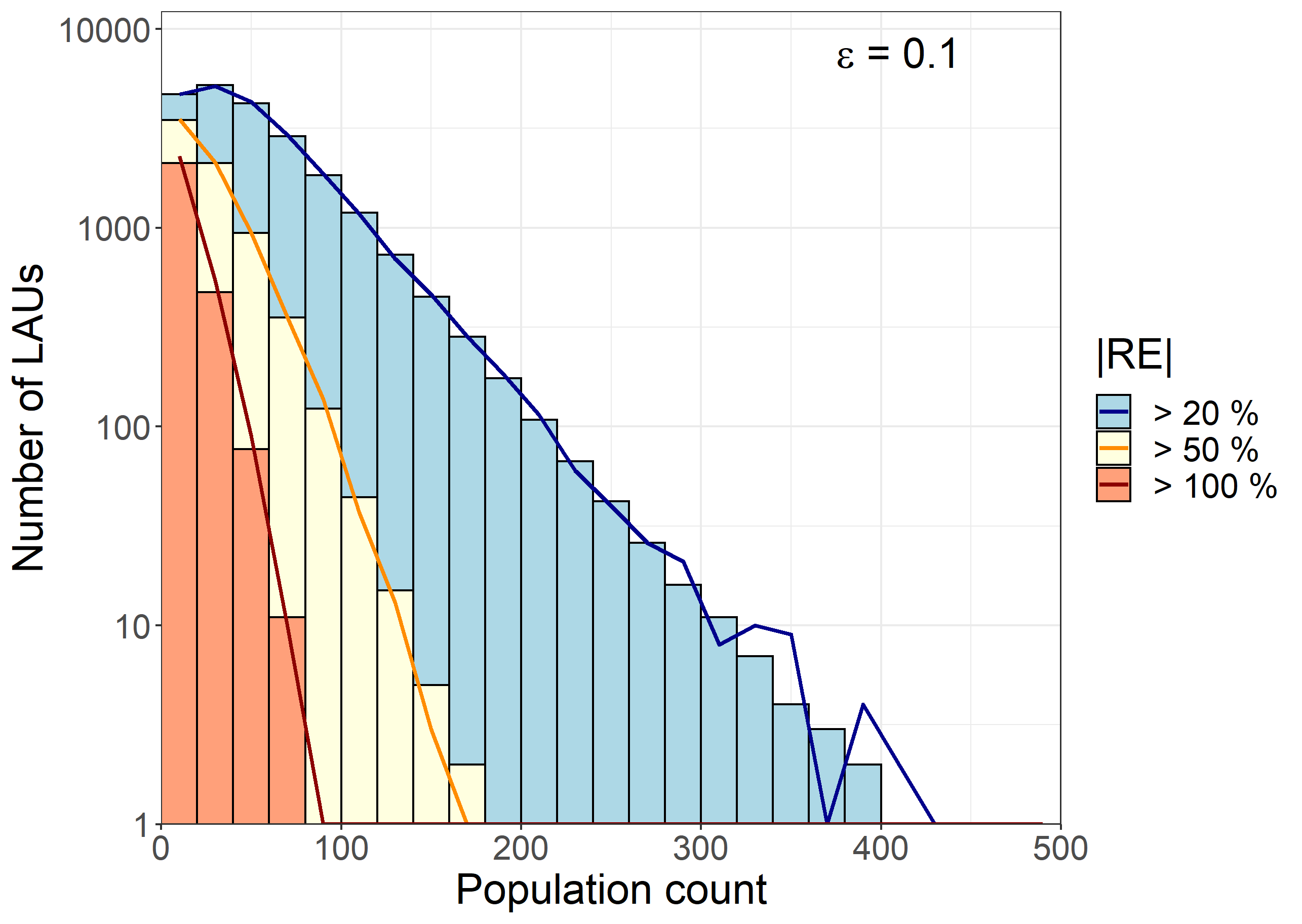}
    \includegraphics[width=0.49\textwidth]{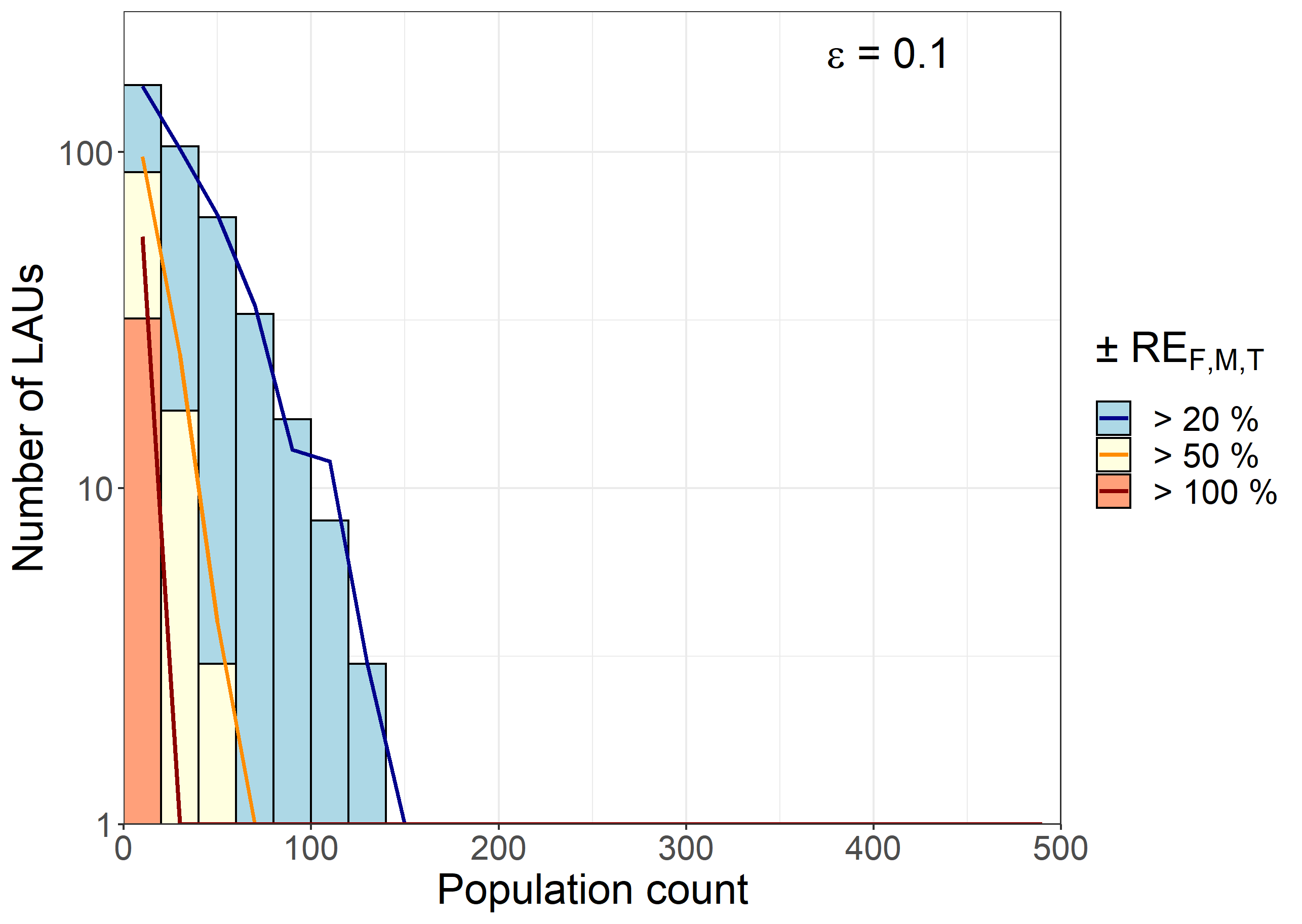}
    \includegraphics[width=0.49\textwidth]{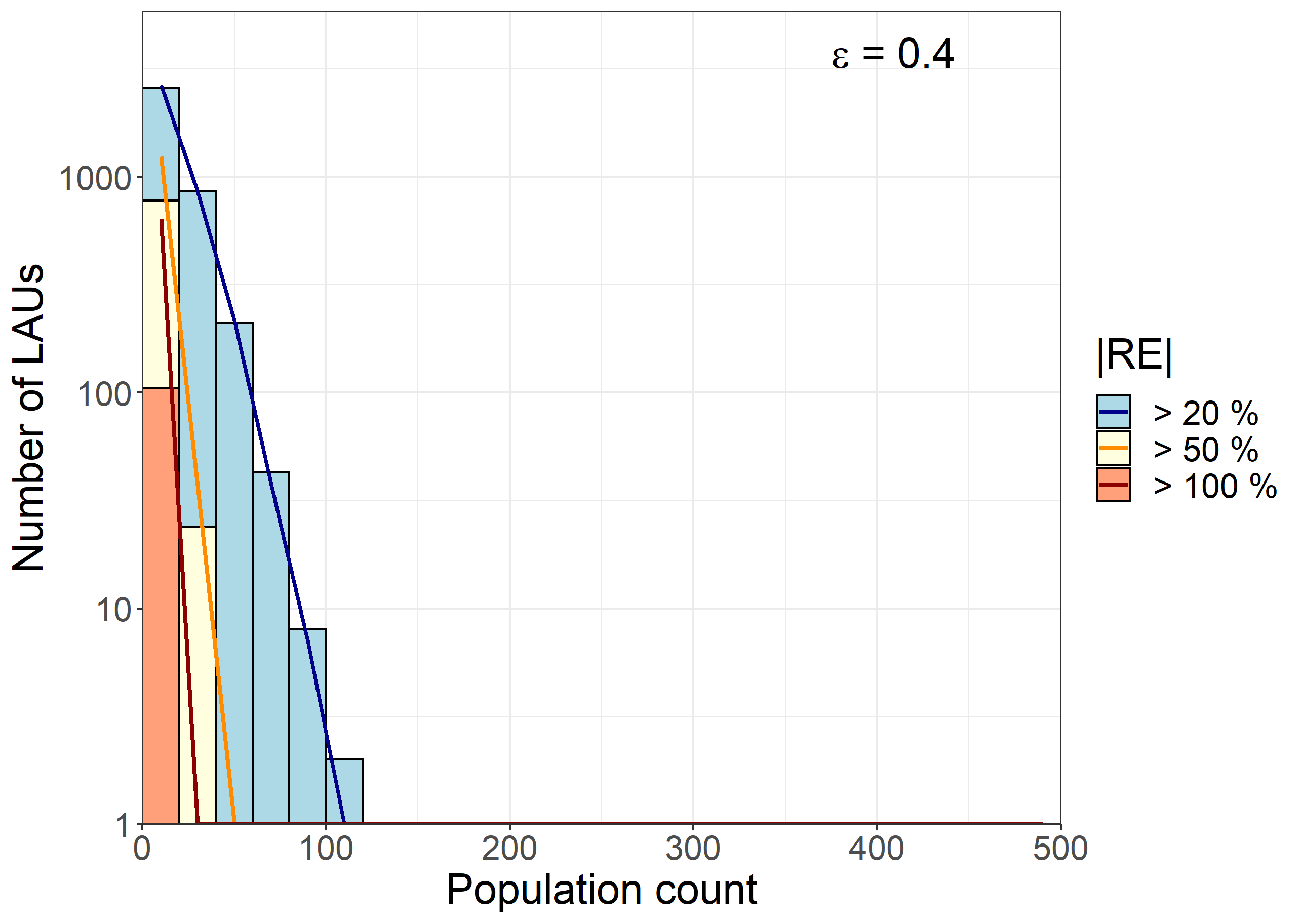}
    \includegraphics[width=0.49\textwidth]{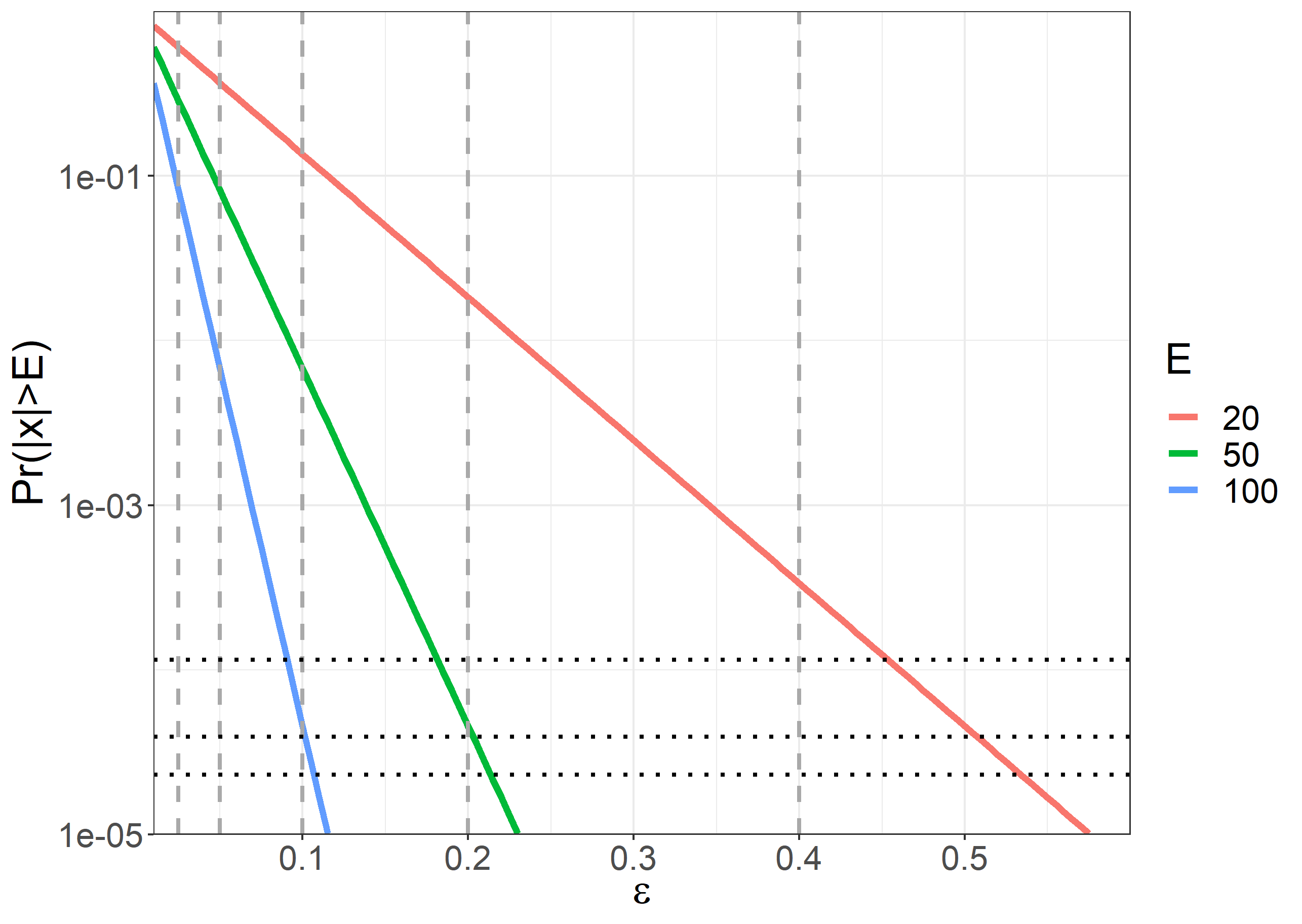}
    \caption{Log-linear estimates for frequencies of relative error (RE) magnitudes exceeding $20\percent$ (blue), $50\percent$ (yellow) and $100\percent$ (orange) occurring in LAU counts, by total count: bins show the analytic estimate obtained from Eq.~\eqref{eq_lap_E}, while lines show the actual distortion frequencies found in the data with noise sampled.\newline
    The rows vary $\eps=0.025$ (top) to $0.1$ (middle) to $0.4$ (bottom). The left column counts only single observations ($F$, $M$ or $T$) that exceed the relative noise magnitude, while the right column counts LAUs where all $F$, $M$ and $T$ exceed the relative noise in the same direction.\newline
    The lower right histogram ($F$, $M$ and $T$ distorted in the same direction for $\eps=0.4$) is almost empty and thus replaced by a plot illustrating Eq.~\eqref{eq_lap_E}: log-linear $\pr(|x|>E)$ as a function of $\eps$ with $E=20$ (orange), 50 (yellow) and 100 (blue).  Vertical dashed lines indicate $\eps$ choices from Eq.~\eqref{eq_us_setup}, while horizontal dotted lines show 1 over the number of LAUs with $T\leq E=20$, 50 or 100.}
    \label{fig_utility_lau_2}
\end{figure}

\textbf{Estimating distortions}\quad The basis for $\eps$-DP noise analysed here is the discrete two-tailed geometric distribution, Eq.~\eqref{eq_geom2}, with an $\eps$ range given in Eq.~\ref{eq_us_setup}.  However, in this $\eps$ range the discrete distribution already converges well to the continuous $\lap(1/\eps)$.  The cumulative inverse distribution function of $\lap(1/\eps)$ can be used to calculate the probability for the noise magnitude $|x|$ to exceed a certain threshold $E$:
\begin{equation}\label{eq_lap_E}
    \pr(|x|>E|\eps) = -\exp\left(\eps E\right).
\end{equation}
This probability is plotted in the lower-right of Fig.~\ref{fig_utility_lau_2} as a function of $\eps$ inside the relevant range, and for $E\in\{20,50,100\}$.  Now Eq.~\eqref{eq_lap_E} can be convoluted with the LAU count distribution (left plot in Fig.~\ref{fig_utility_lau_1}) to estimate how many LAUs in each bin will end up with an output count exceeding a given relative error (RE) magnitude threshold.\footnote{{\it Example:} There are $11\,680$ observations ($F$, $M$, or $T$) in the count bin $(60,80]$.  Choose $\mathrm{RE}=50\percent$ and use Eq.~\eqref{eq_lap_E} with $\eps=0.1$ and $E=50\percent\times 80=40$ (taking the right bin limit $80$ for a conservative estimate) to obtain the probability $\sim 1.83\percent$ that a count in this bin will be distorted by at least $50\percent$.  So at least $1.83\percent\times 11\,680\simeq 214$ observations are expected to be affected.  In the noise-sampled data we find 339 such observations.}
These binned estimates can be tested by actually sampling some noise on the LAU data, and counting occurrences of RE magnitudes above a given threshold.  Fig.~\ref{fig_utility_lau_2} (left column) overlays the estimates with counts found in the noise-sampled data.  Clearly the analytic estimates describe very well the noise-sampled data.

\textbf{Distortions of single counts} Looking now at the actual distortions in the left column of Fig.~\ref{fig_utility_lau_2}, one finds a sizeable dependence on $\eps$, which is not surprising due to the exponential noise scaling with $\eps$.  In fact, noise distortions of single counts in these LAU statistics may be said to become manageable from $\eps>0.4$ (and we do not show the upper end of the range in Eq.~\eqref{eq_us_setup}, $\eps=0.8$ for this reason). However, for $\eps\lesssim 0.1$ there are many LAU counts expected with $|\mathrm{RE}|>50\percent$ or even $>100\percent$:

For instance, with $\eps=0.025$ ($\sqrt{V_\mathrm{table}}=56.6$) there are $1\,648$~observations above~100 affected by $\pm 100\percent$ or more, and still 87~observations above~200 with $\mathrm{RE}\pm 100\percent$ or more.  Recall that every third of these observations describes a total count, and every 6th a total count with $RE<-100\percent$, thus wiping out the whole population of that LAU.  The largest LAU where this happens is Aragnouet, France, with originally 239~residents (now $-7$).  The largest male population that lost all their females is in Bla\v{z}ejov, Czechia, with originally 247~men (now~257) and originally 187~women (now~$-58$), while the largest female population now without men is in Graden, Austria, with originally 252~women (now~237) and originally 230~men (now $-57$).

The situation does improve with $\eps=0.1$ ($\sqrt{V_\mathrm{table}}=14.1$), but we still find $122$~observations above~40 and 11~observations above~60 with $\mathrm{RE}\pm 100\percent$ or more.  The largest depopulated LAU is again in France, M\'elagues with originally 63~residents (now $-9$); the largest male population now without females: Velleguindry-et-Levrecey, France with $M=82\rightarrow 74$ and $F=76\rightarrow -5$; and the largest female population now without men: Lavans-sur-Valouse, France with $M=77\rightarrow -15$ and $F=62\rightarrow 52$.  All these findings are disconcerting in their own right: if the total count is affected so severely and inner tables are adjusted to the new total, entire LAU populations disappear from the census output.  If inner cells are not adjusted, one could still try to estimate a $T\leq 0$ with $F+M$, but additional noise effects may be huge; even worse with estimating $F\leq 0$ by $T-M$ (and $M\leq 0$ respectively).

\textbf{Distortions of entire LAUs} To illustrate that even ad hoc `repair' estimates will not always help, one can count all LAUs where $F$, $M$ {\it and} $T$ are distorted in the same direction (``broadband distortions''); results are shown in the right column of Fig.~\ref{fig_utility_lau_2}.  For $\eps=0.025$ there are 28~LAUs above 40~residents and 4~LAUs above~80 with a broadband distortion $-100\percent$ or more.  The largest such LAU is Landremont, France with $F=61\rightarrow -8$, $M=74\rightarrow -26$ and $T=135\rightarrow -83$.  For $\eps=0.1$, most broadband distortions of $\pm 100\percent$ only occur in the lowest count bin $(0,20]$, but there is one above: this time Spain, Cidam\'on with $F=15\rightarrow -9$, $M=20\rightarrow -1$ and $T=30\rightarrow -17$.  Broadband distortions $\pm20\percent$ still occur for 61~LAUs with 100 or more residents. The largest LAU where this happens is Ellend, Hungary with $F=112\rightarrow 74$, $M=94\rightarrow 65$ and $T=206\rightarrow 158$.  Even distortions around $\pm 20\percent$ may have significant policy effects at local level.

\begin{figure}[t]  
    \centering
    \includegraphics[width=0.6\textwidth]{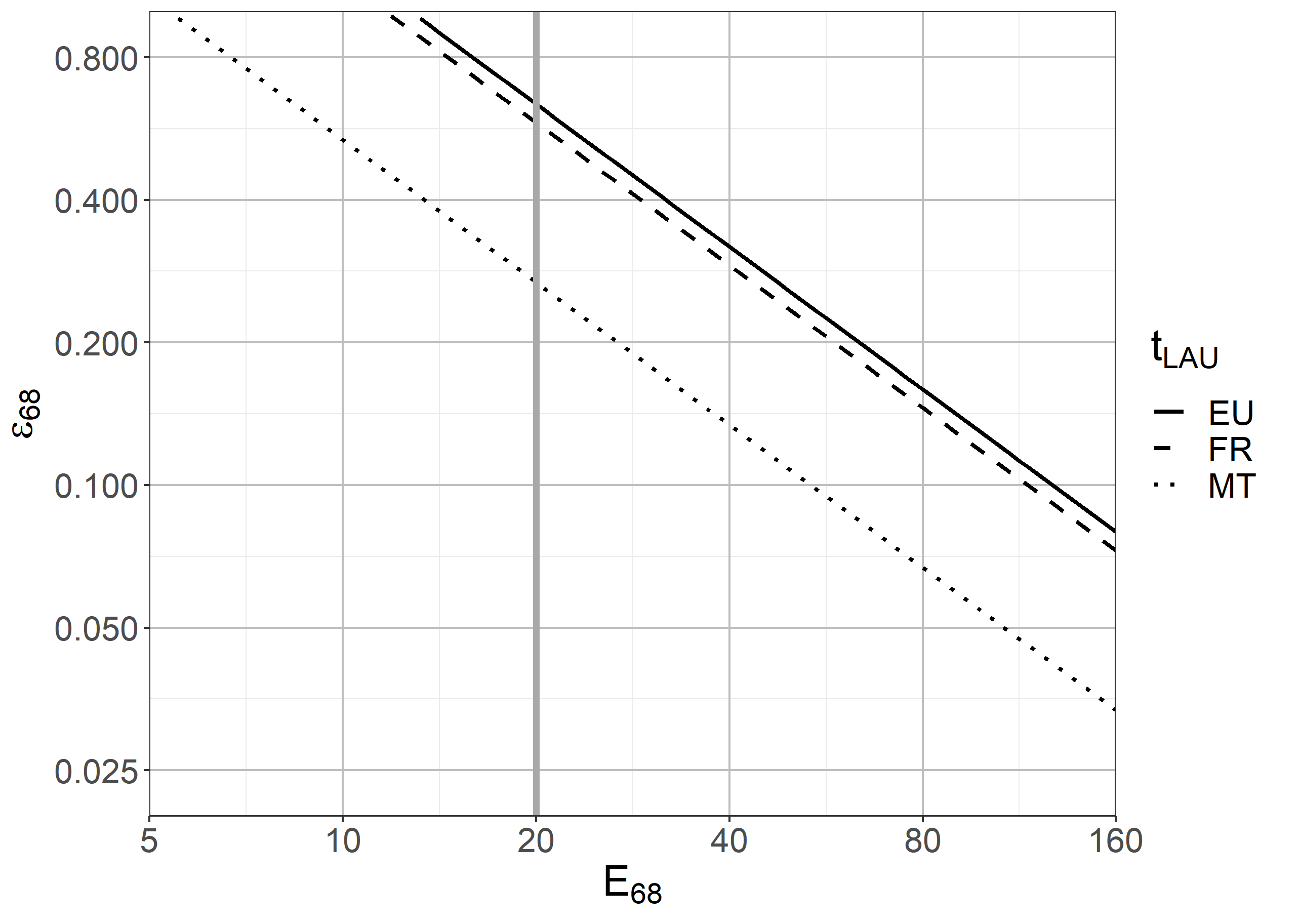}
    \caption{Log-log plot of $\eps_\alpha$ as a function of $E_\alpha$, Eq.~\eqref{eq_dp_utility}, for $\alpha=68\percent$ and three LAU multiplicities: EU-27 with $t_\mathrm{LAU}\simeq 1.1\times 10^5$, France with $t_\mathrm{LAU}\simeq 3.7\times 10^4$ and Malta with $t_\mathrm{LAU}= 68$.}
    \label{fig_utility_lau_3}
\end{figure}

\textbf{Heuristic utility constraint on $\eps$-DP noise}\quad Clearly the results above just paint in vivid colours what unbounded count-level noise size $\sqrt{V}\gtrsim\order(10)$ means for the accuracy of individual counts $\lesssim\order(10^2)$.  Recall that acceptable accuracy of such counts---at least for the small-area output statistics---is a design requirement for most censuses.  Recalling further that any unbounded noise distribution does have a probabilistic bound, Eq.~\eqref{eq_eps_a} in section~\ref{risk_DP}, this can be turned into a utility constraint on $\eps$-DP noise distributions: just require that a given bound $E_\alpha$ is probabilistically tight at confidence level~$\alpha$ for $t$~output counts, which gives
\begin{equation}\label{eq_dp_utility}
    \eps_{\alpha}(E_\alpha|t,\alpha) > \frac{1}{E_\alpha}\log\left( \frac{t}{1-\alpha} \right).
\end{equation}
Within the scope of this paper, a reasonable choice could be $E_\alpha=20$ and $t$ the number of LAUs in an EU Member State.  This would give a (weak) utility guarantee that no LAU total count is distorted $>20$ at $\mathrm{c.l.}=\alpha$.  Fig.~\ref{fig_utility_lau_3} illustrates the implications on count-level $\eps$ values.  In particular, the previous notion is confirmed that individual $\eps\gtrsim 0.6$ (large countries like France, Germany) resp.\ $\eps\gtrsim 0.3$ (small countries like Malta) would have to be spent at least on the small-area outputs to maintain unique census utility.


\subsection{Discussion}\label{utility_discussion}

The simple analysis above has shown that tail effects of unbounded noise distributions, such as strictly $\eps$-DP ones, may cause grave distortions at small geographies.  This starts to kick in severely around count-level $\eps_\mathrm{table}<0.4$ for most countries ($>\order(10^3)$ LAUs).  These results point at similar conclusions as in~\cite{santos2020}: with unbounded noise it is very difficult to maintain a certain minimum utility per individual small area unit, for {\it every} small area unit in the output.

Of course, one could now enter the game of redistributing privacy budgets between geographies or between other statistics, but such fine-tuning is beside the point because it is extremely difficult, on principle, to avoid large tail distortions in an output corner that may turn out to be critical.  Moreover, the risk analysis of section~\ref{risk_av} has shown that, in static outputs of EU 2021 census complexity and without the SPSN principle, count-level $\eps\gtrsim 0.3$ leak into an averaging-vulnerable regime.  This suggests that there is only a narrow band around count-level $\eps\simeq 0.3$ to~$0.4$ that may reduce utility and risk concerns to an acceptable level.

There is an obvious solution that avoids any of these concerns or haggling of budgets between output statistics: do not use unbounded noise for population statistics where geographic granularity is a key element.  This does take strictly $\eps$-DP mechanisms out of the game, but as argued in section~\ref{risk}, this has no practical consequences on risk aspects in static outputs.  If a formal privacy guarantee is desirable for some reason, $\epsdel$-DP mechanisms with risk-driven parametrisations based on truncated noise distributions as \eg in~\cite{rinott2018} are available.  In appropriate setups, such $\epsdel$-DP mechanisms should  perform very similar to utility-driven approaches such as CK, both from a risk and from a utility perspective.

\section{Risk vs.\ utility for upcoming censuses}\label{censuses}

\begin{figure}[t]  
    \centering
    \includegraphics[width=0.49\textwidth]{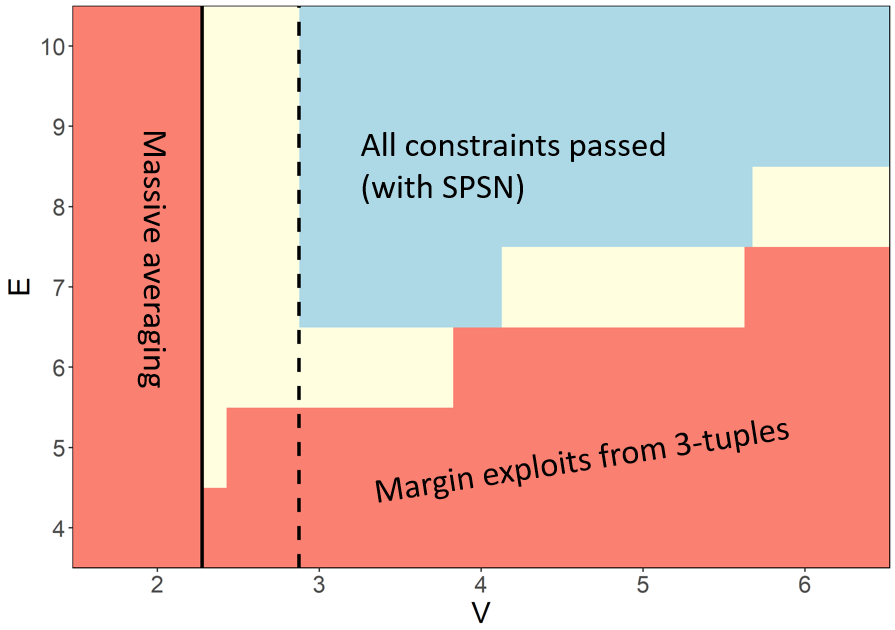}
    \includegraphics[width=0.49\textwidth]{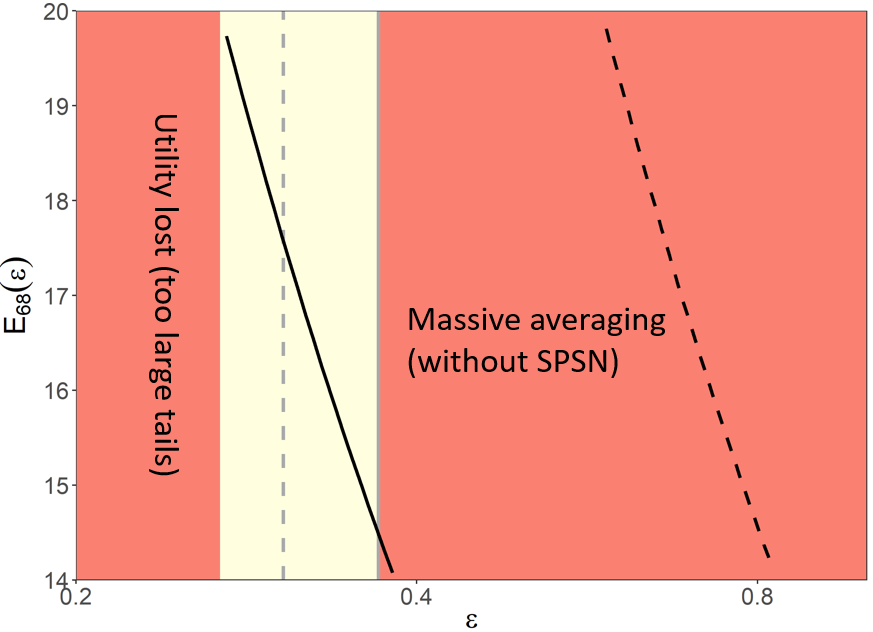}
    \caption{Generic noise parameter space highlighting regions that survive all risk/utility constraints (blue/yellow): the utility-driven generic $V$--$E$ plane (left) and the risk-driven (count-level) $\eps$ range (right).  Yellow regions are not excluded; they rather indicate that such setups may work in certain circumstances, or with slightly relaxed constraints.  Note that the $\eps$ range (right) is a one-parameter space, where a utility constraint is taken from Eq.~\eqref{eq_dp_utility} (note different $E$ scales).  The SPSN principle is assumed to be invoked on the left, but not on the right (DP default).\newline
    On the right, no blue region survives all constraints conservatively: averaging (grey lines showing $\eps$ for smallest $k/t^2$ dashed and for next-smallest solid) and $E_\alpha<20$ at $\alpha=68\percent$ (black lines showing the $E_\alpha(\eps)$ curve for Malta solid and for France dashed).  When relaxing certain constraints (\eg $E_{68}\gtrsim 20$ and/or slight averaging vulnerability), a small yellow band $\eps\in[0.27,0.37]$ remains.}
    \label{fig_censuses}
\end{figure}

In an attempt to integrate the findings of sections~\ref{risk} and~\ref{utility} for the scope of the 2021 EU census, the parameter constraints of Figs.~\ref{fig_risk_BN}, \ref{fig_risk_av} and~\ref{fig_utility_lau_3} can be combined into a global picture of the generic noise parameter space:  Fig.~\ref{fig_censuses} illustrates that utility-driven parametrisations using individual count-level variance $V$ and noise bound~$E$ can be set up within a range that avoids all risk/utility constraints assessed in this paper (\eg $V\simeq 2$ to~3 and $5\lesssim E\lesssim 10$).  On the other hand, risk-driven approaches such as strictly $\eps$-DP mechanisms with unbounded noise are severely constrained by the simultaneous requirements of risk (massive averaging) and utility (small-area accuracy) considerations.  In particular, only a narrow window around individual count-level $\eps\simeq 0.3$ seems to remain with acceptable compromises.

Global constraints as in Fig.~\ref{fig_censuses} do depend on the exact (static) output, but in general such constraints can always be obtained systematically from the static output structure.  This is what makes the risks controllable: if no satisfying parameter setup is found, the output can be curated to relax the constraints.  While $\eps$-DP as a {\it risk measure} may contribute to an assessment of appropriate noise amounts, the flexibility of $\eps$-DP {\it mechanisms} is heavily limited with just a single parameter (the privacy budget).  It is the presence of a second parameter---the noise bound $E$, or $\delta$ in $\epsdel$-DP mechanisms---that adds flexibility to arbitrate between risk and utility constraints.\footnote{For instance, if the noise amount should be increased, this is often possible by tuning $V$ up while keeping $E$ fixed.  In $\eps$-DP mechanisms one can only reduce $\eps$, which often has direct and severe consequences on utility through the noise tails.\label{fn_2-pars}}

In view of recent developments such as the \uscb\ embracing differential privacy for their 2020 census~\cite{abowd2018} and new studies indicating a vulnerability of bounded noise to massive averaging attacks in certain output mechanisms~\cite{ashgar2019}, some targeted remarks are in order:
\begin{itemize}
    \item In most countries, censuses are very costly investments that must be justified by their unique value added for policymakers and researchers. Consequently, user concerns must be taken very seriously, and also SDC methodology must acknowledge key utility constraints.
    \item Section~\ref{risk} has illustrated for typical systematic attacks on static outputs, such as exploiting margins or massive averaging, that strictly $\eps$-DP mechanisms (with unbounded noise) do not offer significant benefits over utility-driven approaches with bounded noise, including $\epsdel$-DP ones.
    \item Section~\ref{utility} has demonstrated that mechanisms based on unbounded noise (incl.\ strictly $\eps$-DP mechanisms) may ruin unique utility features of censuses, such as small-area accuracy.  This is in line with concerns related to the envisaged 2020 U.S.\ census setup~\cite{ruggles2019,santos2020}.
\end{itemize}
Thus in static outputs utility-driven parametrisations, which transparently limit utility loss at the expense of increased efforts to control and communicate disclosure risks, seem more favourable than risk-driven parametrisations, which transparently control disclosure risks at the expense of potentially uncontrolled utility loss.  In particular with static outputs, strictly $\eps$-DP mechanisms waive increased utility guarantees provided by tight noise bounds without need.

\section{Conclusions}\label{conclusion}

While traditional statistical disclosure control (SDC) methods in population statistics mainly focused on small counts at high risk of direct re-identification (\eg\ suppressing those, etc.), powerful theoretical results have meanwhile shown that entire microdata data\-bases can often be reconstructed accurately from too detailed output tables~\cite{dinur2003}, thus exposing rare records even if small output counts were treated.  These results suggest that random noise methods are the most effective counter-measure, where the amount of noise should scale with output detail.  This scaling rule implies a first important notion: flexible output mechanisms (where the complexity is not fixed a priori) require some kind of noise scaling and are thus much harder to realise within reasonable risk and utility constraints.  On the other hand, static output mechanisms (pre-fixed complexity) allow for a diligent curation, including controlling risks and assessing risk/utility trade-off to fix a static noise amount.  Unless imposed by external constraints, a move from a static to a flexible output mechanism should be considered only with utmost care.

Differential privacy (DP) is a useful concept to quantify risk irrespective of a particular output scenario, and hence to compare risk levels consistently between various SDC approaches~\cite{dwork2006,dwork2014}.  DP risk measures may thus contribute to a broadly based SDC assessment.  Moreover, DP provides for automatic noise scaling with output complexity, as required by flexible output mechanisms.  However, this paper suggests that the complexity scaling of DP noise levels is over-protective for increasingly complex outputs, so DP inferences on absolute noise levels should be handled with care, especially with complex static outputs.

The paper makes a clear separation between DP risk measures and DP output mechanisms, where the latter may give strict $\eps$-DP or relaxed $\epsdel$-DP privacy guarantees with $\eps$ the total privacy budget spent on the entire output.  However, strictly $\eps$-DP mechanisms must employ unbounded random noise distributions, while relaxed $\epsdel$-DP or not manifestly DP mechanisms can have bounded distributions.  It is shown that in static output scenarios, typical generic risks such as margin exploits and massive averaging are controllable with bounded noise, $\epsdel$-DP or not.  Conversely, the unbounded noise of strictly $\eps$-DP mechanisms may lead to severe utility damage when the noise amount is tuned up to evade averaging risks.  More generally, the fact that $\eps$-DP mechanisms only have a single parameter costs a lot of flexibility.

Censuses are big national investments for comparably narrow purposes, not necessarily to answer any question any user may have on any characteristics of any sub-population.  This suggests a static output mechanism with a utility-driven parametrisation, which allows to maximise utility within purpose scope while controlling risks carefully.  Finally, if particular SDC mechanisms jeopardise unique census features, they are bluntly unfit for the purpose.  For the scope of the 2021 EU census round, this paper finds noise methods recommended by the European Statistical System~\cite{essnet2017}, including bounded noise from the cell key method, suitable to protect outputs in a controlled way.  The generic parameter space (noise variance and noise bound) is constrained by different risk or utility requirements, but various setups remain feasible.  Such setups can obtain a relaxed $\epsdel$-DP guarantee, if needed.  On the other hand, strictly $\eps$-DP mechanisms are severely constrained, with only a small parameter window remaining for a possibly acceptable compromise.  It seems strict $\eps$-DP guarantees are overpriced (in utility) at least for census-like scenarios.

\section*{Acknowledgments}

The author would like to thank Peter-Paul de Wolf, Tobias Enderle and Fabio Ricciato for draft reading and very useful exchanges.

\appendix

\section{ESS recommendations for harmonised 2021 EU census protection}\label{a_eu_census_sdc}

Within the legal framework for EU censuses\footnote{Regulation (EC) No 763/2008 of the European Parliament and of the Council of 9 July 2008 on population and housing censuses (\href{https://eur-lex.europa.eu/legal-content/EN/TXT/?qid=1474460202679&uri=CELEX:32008R0763}{OJ~L~218, 13.8.2008, p.~14}).\label{fn_cfr}},
statistical confidentiality and SDC measures are under the responsibility of the national statistical authorities of the Member States, so that confidential data must not be submitted to Eurostat.\footnote{See Articles 2(5) and 4(3) of footnote~\ref{fn_cir-2}.\label{fn_c-flag}}
This means that Eurostat will not receive, maintain or process any confidential data (or personal data in GDPR sense) related to the upcoming 2021 EU census round.  This has lead to various ESS level projects facilitated by Eurostat and aimed at developing common ESS recommendations for harmonised methods~\cite{essnet2017} and tools~\cite{essnet2019} to protect census outputs at the national level.  These recommendations include the cell key (CK) method as a particular mechanism implementing bounded noise~\cite{meindl2019}.

During the project work, considerable efforts were made to provide more flexible and accessible SDC tools~\cite{essnet2019a} but also to assess disclosure risks vs.\ utility and improve the general methodology~\cite{giessing2016,enderle2018,enderle2020}.  Also the findings of this paper (see section~\ref{censuses}) suggest that the ESS recommended CK method stands as a good practice, if applied correctly and consistently, featuring in particular superior utility properties in various output scenarios with still small and controllable disclosure risks.

\textbf{Small count threshold parameter}\quad Note that the CK method formally takes a third parameter (in addition to variance $V$ and noise bound $E$), namely a threshold $j_s$ for the smallest non-zero count that may occur in the output (\ie no output count will be $>0$ and $\leq j_s$).  This may be desirable for some national output curators, \eg for historical or cultural reasons.  The CK method implements $j_s$ in a consistent manner without introducing biases, as opposed to na\"ive noise truncation $\geq 0$~\cite{ghosh2012,petti2019}.  However, recent results indicate that a $j_s>0$ actually increases risks and may loosen a corresponding $\epsdel$-DP guarantee~\cite{rinott2018}.  Having a $j_s>0$ or not does not affect any of the arguments of this paper significantly, so we just fixed $j_s=0$ for all analyses presented.

\textbf{The European census grid}\quad is a notable new output from the 2021 EU census round.\footnote{Commission Implementing Regulation (EU) 2018/1799 of 21 November 2018 on the establishment of a temporary direct statistical action for the dissemination of selected topics of the 2021 population and housing census geocoded to a $1\,\mathrm{km}^2$ grid (\href{https://eur-lex.europa.eu/legal-content/EN/TXT/?uri=uriserv:OJ.L_.2018.296.01.0019.01.ENG}{OJ~L~296, 22.11.2018, p.~19}).\label{fn_cir-4}}
Several key census indicators will be published for each cell of a pan-European $1\,\mathrm{km}^2$ grid, which poses specific new disclosure risks (see~\cite{bach2018} for an overview).  However, main risks relate to the possible combination of different non-nested small areas (\eg\cite{costemalle2019}), where again random noise is considered effective~\cite{essnet2017}.  For the scope of this paper the grid output is not important: the additional number of 3-tuples is negligible,\footref{fn_3-tuples_grid} and it does not add any redundancy to any of the other output statistics because the grid does not overlap with administrative geographic breakdowns and it has no `total'.

\section{Massive averaging in static outputs}\label{a_av}

\subsection{Output structure}\label{a_av_output}

In static output scenarios, all output statistics are fixed in advance.  In census-like scenarios, these are typically contingency tables which cross-tabulate several categorical variables with pre-fixed, finite sets of values (variable breakdowns).  The tables may contain margins, \ie there may be a hierarchical structure inside the variable breakdowns (some values are contained within other values).  In particular, let each variable breakdown contain a value ``total'', which denotes the union of all other (``internal'') values of that variable.  Finally, a table cell is a distinct combination of variable values, where the cell value (without noise injection) is the number of microdata records that are characterised by the combination of variable values defining the cell.

Let $A$ denote a variable breakdown represented by a finite set of discrete values $a\in A$, where generally $A=\{\cdot,\tot\}$ and $|A|$ the cardinality of $A$.  Let $\mathbf{A}$ denote an $m$-tuple of variable breakdowns $\{A_i\}_m$, so that $|\mathbf{A}|=m$, and let $\mathbf{a}\in\mathbf{A}$ denote an $m$-vector of variable values $\{a_i\}_m$ with $a_i\in A_i\;\forall\; i\in\{1\cdots m\}$.  Then $T_\mathbf{A}$ represents the $m$-dimensional cross-tabulation of $\mathbf{A}$ (a table), and $T_\mathbf{a}$ a single table cell thereof.  Furthermore, $T_{\mathbf{A^\prime},\mathbf{a^{\prime\prime}}}$ represents the $|\mathbf{A^\prime}|$-dimensional sub-table of $T_\mathbf{A}$ obtained by fixing variables $\mathbf{A^{\prime\prime}}$ to values $\mathbf{a^{\prime\prime}}$, and $\mathbf{A^\prime}\,\dot\cup\,\mathbf{A^{\prime\prime}}=\mathbf{A}$. Now suppress all $a_i=\tot$, such that $T_\mathbf{A^\prime}\subset T_\mathbf{A}$ represents the $|\mathbf{A^\prime}|$-dimensional marginal table for $\mathbf{a^{\prime\prime}}=\{\textrm{total}\}_{|\mathbf{A^{\prime\prime}}|}$.  Thus the table $T_\mathbf{A}$ consists of
\begin{equation*}
    |\mathcal{P}(\mathbf{A})|=\sum_{i=0}^m {m\choose i} = 2^m
\end{equation*}
sub-tables, including itself, all marginal tables and the total margin $T$, where $\mathcal{P}(\mathbf{A})$ is the power set of $\mathbf{A}$.

On the other hand, we can sum all table cells characterised by $\mathbf{a}=\{\mathbf{a}_{m-1},a_m\}$ over all values $a_m\in A_m$ to obtain an independent redundant representation (IRR) of the marginal cell $T_{\mathbf{a}_{m-1}}$, or in short-hand notation
\begin{equation}\label{eq_red}
    \sum_{i=1}^{|A_m|} T_{\{\mathbf{a}_{m-1},a_{m,i}\}}\coloneqq T_{\mathbf{a}_{m-1}}{}^{\mathbf{A}_m}
\end{equation}
and $T_{\mathbf{A}_{m-1}}{}^{\mathbf{A}_m}$ the respective marginal table.  Thus, generally, $T_\mathbf{A^\prime}{}^{\overline{\mathbf{A^\prime}}}$ is an IRR of $T_\mathbf{A^\prime}$ for any possible disjoint partition of $\mathbf{A}=\{\mathbf{A^\prime},\overline{\mathbf{A^\prime}}\}$.  Then the set of all IRRs of a target cell $T_\mathbf{a^\prime}$ (or entire target table $T_{\mathbf{A}^\prime}$) is given by $\mathcal{P}(\overline{\mathbf{A^\prime}})$ as $\{T_\mathbf{a^\prime}{}^{\mathbf{A}_i}\}_{\mathbf{A}_i\in\mathcal{P}(\overline{\mathbf{A^\prime}})}$, or short $T_\mathbf{a^\prime}{}^{\mathcal{P}(\overline{\mathbf{A^\prime}})}$.

{\it Example:} Table~9.2 of the 2021 EU census programme\footref{fn_cir-2} 
is defined as
\begin{equation}\label{eq_hc9.2}
    \textrm{9.2:}\quad \textrm{GEO.M} \times \textrm{SEX} \times \textrm{AGE.M} \times \textrm{YAE.H},
\end{equation}
where GEO.M is the geographic breakdown of medium detail (NUTS~3), $\textrm{SEX}=\{\textrm{F},\textrm{M},\textrm{total}\}$ is the sex breakdown, AGE is the medium age breakdown (5-year bands) and YAE.H is the highly detailed breakdown of year of arrival in the reporting country (by single years).  This table is expressed as $T_{\mathbf{A}}$, where $\mathbf{A}=\{\textrm{GEO.M}, \textrm{SEX}, \textrm{AGE.M}, \textrm{YAE.H}\}$, and $|\mathbf{A}|=4$ so there are 16 possible sub-tables or subsets of $\mathbf{A}$.  For instance, let $\mathbf{A^\prime}=\{\textrm{GEO.M}, \textrm{SEX}\}$.  Then $\overline{\mathbf{A^\prime}}=\{\textrm{AGE.M}, \textrm{YAE.H}\}$ and
\begin{equation*}
    \mathcal{P}\left(\overline{\mathbf{A^\prime}}\right)=\{\emptyset,\{\textrm{AGE.M}\},\{\textrm{YAE.H}\},\{\textrm{AGE.M}, \textrm{YAE.H}\}\}.
\end{equation*}
Thus, we find for the table of interest $T_{\{\textrm{GEO.M}, \textrm{SEX}\}}$ four IRRs inside $T_{\mathbf{A}}$: $T_{\{\textrm{GEO.M}, \textrm{SEX}\}}$ itself (the trivial marginal table obtained from $\emptyset\in\mathcal{P}(\overline{\mathbf{A^\prime}})$) as well as $T_{\{\textrm{GEO.M}, \textrm{SEX}\}}{}^{\{\textrm{AGE.M}\}}$, $T_{\{\textrm{GEO.M}, \textrm{SEX}\}}{}^{\{\textrm{YAE.H}\}}$ and $T_{\{\textrm{GEO.M}, \textrm{SEX}\}}{}^{\{\textrm{AGE.M}, \textrm{YAE.H}\}}$.

\subsection{Averaging attacks}\label{a_av_averaging}

Let the static output consist of $M$ predefined tables $\cup_{I\in\{1\cdots M\}} TI_{\mathcal{P}(\mathbf{A}_I)}$.  This fixes the entire universe of independent output counts, or table cells before publication, including all hierarchical constraints and redundancies outlined in section~\ref{a_av_output}.  It is thus possible to identify, for any given target count $TI_{\mathbf{a}}$ inside the universe, all IRRs contained in the universe, \ie all $TJ_{\mathbf{a}}{}^{\mathcal{P}(\overline{\mathbf{A}_J})}$ from all $J\in\{1\cdots M\}$ with $\mathbf{A}\subset\mathbf{A}_J$ and $\{\mathbf{A},\overline{\mathbf{A}_J}\}=\mathbf{A}_J$.  The corresponding set is
\begin{equation*}
    \mathcal{T}_\mathbf{a}(\mathbf{A})\coloneqq \cup_{J\in\{1\cdots M\}\;\textrm{where}\;\mathbf{A}\subset\mathbf{A}_J}TJ_\mathbf{a}{}^{\mathcal{P}(\overline{\mathbf{A}_J})},
\end{equation*}
A global constraint using all available redundant representations of the target count is thus given by
\begin{equation}\label{eq_red_global}
    TI_{\mathbf{a}} = \frac{1}{|\mathcal{T}_\mathbf{a}(\mathbf{A})|}\sum_{i=1}^{|\mathcal{T}_\mathbf{a}(\mathbf{A})|} \mathcal{T}_\mathbf{a}(\mathbf{A})_i.
\end{equation}

When the output is protected by noise injection as described in section~\ref{pre}, each independent output cell will get independent noise from a given distribution with variance $V$.  Note that the SPSN principle introduced in section~\ref{pre_output} leads to {\it dependent} noise on all $TI_\mathbf{A}$ and $TJ_\mathbf{A}$ with $I\neq J$ and $\mathbf{A}\subseteq\mathbf{A}_I\cap\mathbf{A}_J$.  Thus, with SPSN we can drop table indices $I$ so that the entire output universe is just $\cup_{I\in\{1\cdots M\}} T_{\mathcal{P}(\mathbf{A}_I)}$.  Defining a set that consists only of unique variable combinations disjoint from $\mathbf{A}$
\begin{equation*}
    \mathcal{U}(\mathbf{A})\coloneqq \cup_{I\in\{1\cdots M\}\;\textrm{where}\;\mathbf{A}\subset\mathbf{A}_I}\mathcal{P}(\overline{\mathbf{A}_I}),
\end{equation*}
the complete set of IRRs of $T_\mathbf{a}$ is just $T_{\mathbf{a}}{}^{\mathcal{U}(\mathbf{A})}$, and Eq.~\eqref{eq_red_global} reduces to
\begin{equation}\label{eq_red_unique}
    T_{\mathbf{a}} = \frac{1}{|\mathcal{U}(\mathbf{A})|}\sum_{i=1}^{|\mathcal{U}(\mathbf{A})|} T_{\mathbf{a}}{}^{\mathcal{U}(\mathbf{A})_i}.
\end{equation}
This does {\it not} hold for noise mechanisms ignoring the SPSN principle (such as generic DP mechanisms), where independent noise is drawn for any $TI_{\mathbf{A}}$ and $TJ_{\mathbf{A}}$ with $I\neq J$, even though $\mathbf{A}$ is fixed, and Eq.~\eqref{eq_red_global} holds.  Therefore, depending whether the principle is enforced or not, Eq.~\eqref{eq_red_global} or Eq.~\eqref{eq_red_unique} can be used to average over all independent representations of the target count available in the output to obtain an estimate $\widetilde{T_\mathbf{a}}$.  Note that in a static output scenario, no other independent representations of the target can be generated by the user: the amount of independent noise is predefined by the complexity of $\{TI_{\mathbf{A}_I}\}_{I\in\{1\cdots M\}}$ resp.\ $\{T_{\mathbf{A}_I}\}_{I\in\{1\cdots M\}}$.\footnote{This is significantly different from the scenario in~\cite{ashgar2019}, where the user can submit many queries asking for {\it custom} bi-partitions of any available variable; in our nomenclature this is equivalent to custom-defining $A=\{a_1,a_2\}$, which is impossible by assumption of this section (static output).\label{fn_bi-partitions}}

\subsection{Disclosure risks}\label{a_av_risk}

Eq.~\eqref{eq_cheb} relates the overall success probability of averaging the target count correctly to the (constant) variance~$V$ of the noise applied to each single independent output, where $k$ is the number of independent outputs being summed and $t$ is the number of independent representations of $T_\mathbf{a}$:
\begin{equation}\label{eq_k_t_global}
    k_\mathbf{a} = \sum_{I\in\{1\cdots M\}\;\textrm{where}\;\mathbf{A}\subset\mathbf{A}_I} \sum_{i=1}^{|\mathcal{P}(\overline{\mathbf{A}_I})|}\prod_{j=1}^{|\mathcal{P}(\overline{\mathbf{A}_I})_i|} \left|\mathcal{P}(\overline{\mathbf{A}_I})_{ij}\right|
    \quad\textrm{and}\quad
    t_\mathbf{a} = |\mathcal{T}_\mathbf{a}(\mathbf{A})|
\end{equation}
resp.
\begin{equation}\label{eq_k_t_unique}
    k_\mathbf{a} = \sum_{i=1}^{|\mathcal{U}(\mathbf{A})|}\prod_{j=1}^{|\mathcal{U}(\mathbf{A})_i|} \left|\mathcal{U}(\mathbf{A})_{ij}\right|
    \quad\textrm{and}\quad
    t_\mathbf{a} = |\mathcal{U}(\mathbf{A})|.
\end{equation}
However, Eq.~\eqref{eq_cheb} gives a {\it lower} limit on the averaging success probability, whereas an {\it upper} limit would be required from a protection point of view.  Therefore, to assess averaging risks conservatively we continue the analysis by postulating a Gaussian noise distribution.  This has the advantage that the expected distribution of the cumulated noise on $\widetilde{T_\mathbf{a}}$ is again Gaussian with variance
\begin{equation*}\label{eq_gaussian_var}
    \var(\widetilde{T_\mathbf{a}})=\frac{k_\mathbf{a}V}{t_\mathbf{a}^2},
\end{equation*}
so that the averaging success probability $\alpha_\mathbf{a}=\pr(|\widetilde{T_\mathbf{a}}-T_\mathbf{a}|<0.5)$ can be calculated exactly.  This approximation performs well \eg in CK noise
scenarios.
\footnote{
Cf.\ footnote~\ref{fn_CK_EU2021-census}; in fact it is easy to convince oneself empirically that the approximation holds well, by generating some dummy samples of CK noise and testing the Gaussian model.  For instance, in 1000 samples with $V=2$, $k=1000$ and $t=100$ (thus $k/t^2=0.1$) we found 741 ($E=10$) resp.\ 725 ($E=5$) samples with averaged noise magnitude $<0.5$, while the Gaussian expectation from Eq.~\eqref{eq_gaussian_var} is $73.6\percent$.\label{fn_ck_gauss_test}}

Eqs.~\eqref{eq_cheb} and~\eqref{eq_gaussian_var} indicate that the ratio $k_\mathbf{a}/t_\mathbf{a}^2$ is a suitable measure of averaging risk, because it fixes $\alpha_\mathbf{a}$ for given~$V$.  However, generally the average over all available IRRs as in Eq.~\eqref{eq_red_global} or Eq.~\eqref{eq_red_unique} does not give the lowest (\ie most risky) $k_\mathbf{a}/t_\mathbf{a}^2$, because some IRRs require very many internal table cells to be added, thus increasing $k_\mathbf{a}$ and the corresponding summed noise amount disproportionately.  This can be accounted for with a simple optimisation:
\begin{enumerate}
    \item Sort all available IRRs from low to high individual $k_\mathbf{a}$ (fixing $I$ and $i$ in Eq.~\eqref{eq_k_t_global} resp.\ $i$ in Eq.~\eqref{eq_k_t_unique} and carrying out the product).
    \item Start averaging iteratively, including new IRRs with increasing $k_\mathbf{a}$ one at a time, as long as the aggregate $k_\mathbf{a}/t_\mathbf{a}^2$ of the average decreases.
    \item The first time the aggregate $k_\mathbf{a}/t_\mathbf{a}^2$ increases, discard the last IRR added and return the previous average as the optimal (\ie most risky) one.
\end{enumerate}

The risk of disclosure from averaging any target count $T_\mathbf{a}$ is thus fixed by the static output complexity $\{TI_{\mathbf{A}_I}\}_{I\in\{1\cdots M\}}$ resp.\ $\{T_{\mathbf{A}_I}\}_{I\in\{1\cdots M\}}$, as well as constant noise variance $V$.  This means an output curator can control for it by either reducing output complexity or increasing $V$.
\begin{table}[t]
\centering
\begin{tabular}{ c||c|c|c||c|c|c| }
    & \multicolumn{3}{c||}{With SPSN} & \multicolumn{3}{c|}{Without SPSN} \\
    \# & MS & $T_\mathbf{A}$ & $k/t^2$ & MS & $T_\mathbf{A}$ & $k/t^2$ \\
    \hline
    1 & LU  & AGE.M & 0.0867 & all & total       & 0.0118 \\
    2 & CY  & AGE.M & 0.0884 & all & GEO.L       & 0.0170 \\
    3 & MT  & AGE.M & 0.0916 & all & AGE.M       & 0.0170 \\
    4 & EE  & AGE.M & 0.108  & all & SEX         & 0.0234 \\
    5 & all & total & 0.112  & all & GEO.L AGE.M & 0.0237 \\
    \hline
\end{tabular}
\caption{The top five smallest optimised $k/t^2$ values found for EU Member States (MS), with and without SPSN.}
\label{tab_risk_BN}
\end{table}

\subsection{Results for the 2021 EU census scenario}\label{a_av_census}

\begin{figure}[t]
    \centering
    \includegraphics[width=0.49\textwidth]{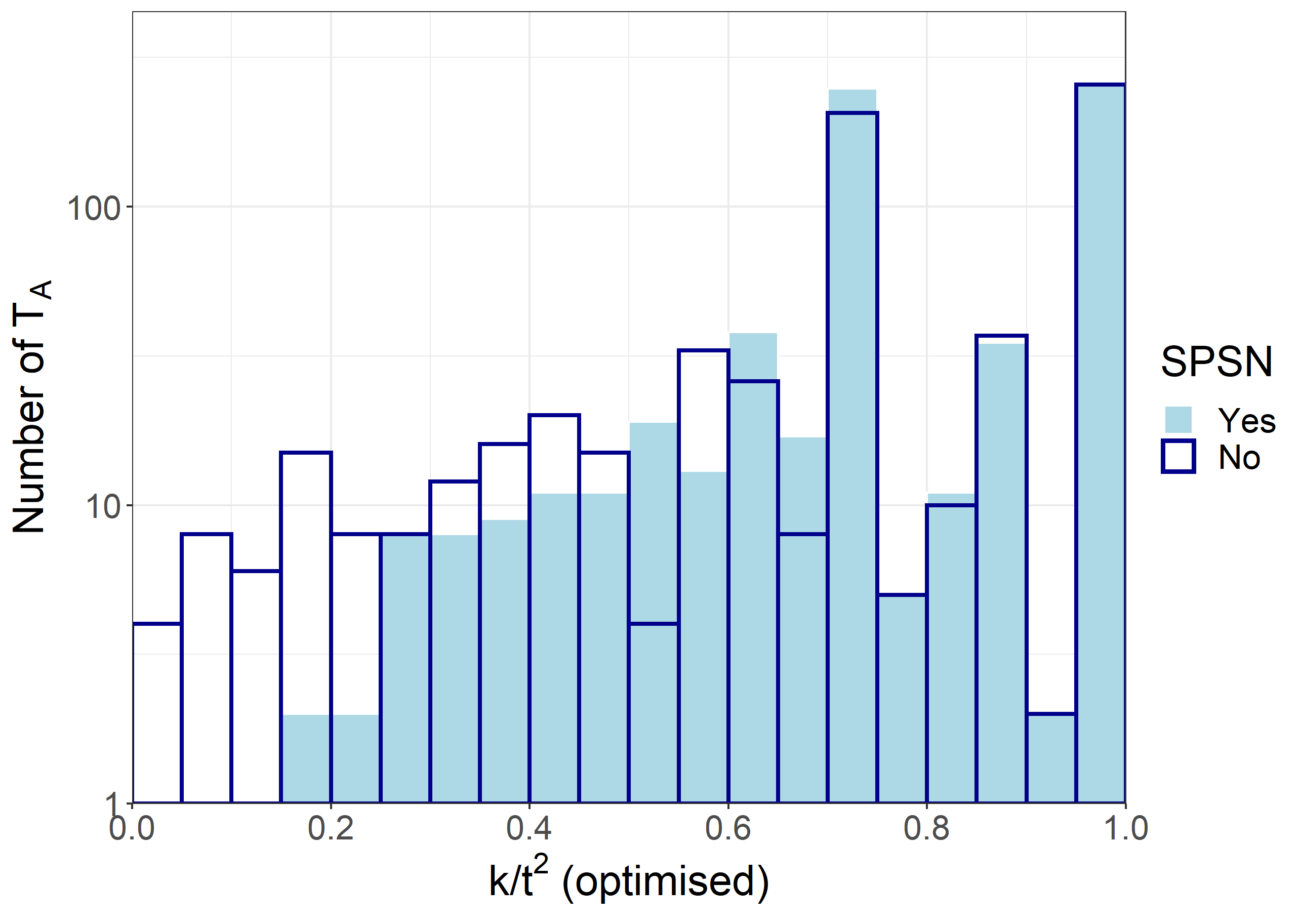}
    \includegraphics[width=0.49\textwidth]{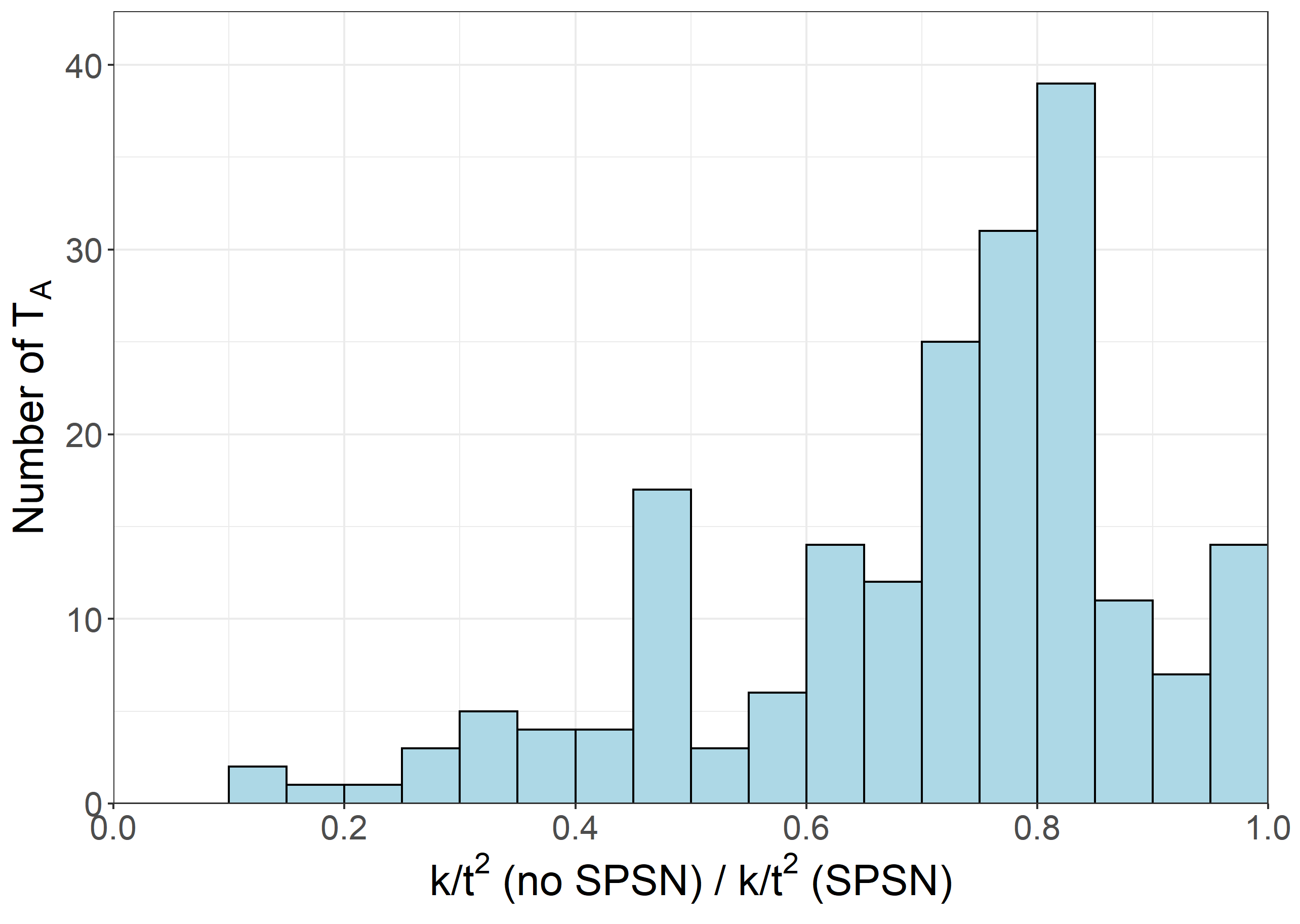}
    \caption{Histograms of the $k/t^2$ distribution across all $T_\mathbf{A}$ without geographic dependence in their averages, with and without SPSN (left), and the ratio of $k/t^2$ with and without SPSN for each of the $T_\mathbf{A}$ (right, ratio of 1 suppressed).}
    \label{fig_risk_BN_2}
\end{figure}

The 2021 EU census programme (\ie output table set) consists of $M=103$ three to six-dimensional tables\footref{fn_cir-2} cross-tabulating counts of natural persons by 32~different variable breakdowns.\footref{fn_cir-2}  While $t_\mathbf{a}$ is fixed for every $\mathbf{a}$ by the table set, for any target cell $T_\mathbf{a}$ without a geographic attribute $k_\mathbf{a}$ generally depends on one or more geographic breakdowns and thus on the reporting country, \cf Table~\ref{tab_risk_BN_2}.  As expected, Table~\ref{tab_risk_BN} shows that the smallest $k/t^2$ values (optimised as described in annex~\ref{a_av_risk}) are found for the smallest countries, where geographic margins contributing to the independent representations have the smallest $k$~weights.  Furthermore, Fig.~\ref{fig_risk_BN_2} shows the distribution of $k/t^2$ across all available $T_\mathbf{A}$ whose IRRs do not depend on any geographic breakdown (mostly where $\mathbf{A}$ contains itself a geographic breakdown, which is never crossed with another geographic breakdown in any of the output tables), \ie those $k/t^2$ that are equally valid for all reporting countries.

Note the difference of almost an order of magnitude in Table~\ref{tab_risk_BN} between the smallest $k/t^2$ when enforcing SPSN, and the smallest $k/t^2$ when ignoring it. Also in the distribution in Fig.~\ref{fig_risk_BN_2}, a sizeable share of the output statistics tends to have smaller $k/t^2$ without SPSN than with it.\footnote{Among the two very frequent $k/t^2$ values apparent in Fig.~\ref{fig_risk_BN_2}, $k/t^2=0.75\equiv 3/4$ stems from IRR sets that only contain the SEX breakdown ($k=2$) and the trivial margin ($k=1$) with total $k=3$ and $t=2$, which often happens for $T_\mathbf{A}$ representing very rare breakdowns. Similarly, $k/t^2=1$ reflects all the uniquely occurring $T_\mathbf{A}$, including all the internal cells of full tables $T_{\mathbf{A}_I}$, which never occur elsewhere as a margin so that their IRRs consist only of themselves ($k=1$ and $t=1$); \cf footnote~\ref{fn_IRR_example}.\label{fn_k_t2_artefacts}}
While the difference is not too big for the majority of target $T_\mathbf{A}$, a systematic attack could first remove the noise from the most vulnerable $T_\mathbf{A}$ and then use those to reduce successively reduce the noise on subsequent $T_\mathbf{A}$ averages. Hence a conservative approach would fix $V$ such that {\it every} $T_\mathbf{A}$ (even with the smallest $k/t^2$) is sufficiently unlikely to be averaged correctly.  Section~\ref{risk_av} discusses implications of this approach on generic noise parameter ranges when protecting 2021 EU census outputs.

\bibliographystyle{unsrt}
\bibliography{references}

\begin{thebibliography}{10}

\bibitem{dinur2003}
Irit Dinur and Kobbi Nissim.
\newblock Revealing information while preserving privacy.
\newblock In {\em Proceedings of the Twenty-second ACM SIGMOD-SIGACT-SIGART
  Symposium on Principles of Database Systems}, pages 202--210, 01 2003.

\bibitem{dwork2006}
Cynthia Dwork, Frank McSherry, Kobbi Nissim, and Adam Smith.
\newblock Calibrating noise to sensitivity in private data analysis.
\newblock In Shai Halevi and Tal Rabin, editors, {\em Theory of Cryptography},
  pages 265--284, Berlin, Heidelberg, 2006. Springer Berlin Heidelberg.

\bibitem{dwork2006a}
Cynthia Dwork, Krishnaram Kenthapadi, Frank McSherry, Ilya Mironov, and Moni
  Naor.
\newblock Our data, ourselves: Privacy via distributed noise generation.
\newblock In {\em Annual International Conference on the Theory and
  Applications of Cryptographic Techniques}, pages 486--503. Springer, 2006.

\bibitem{machanavajjhala2008}
Ashwin Machanavajjhala, Daniel Kifer, John Abowd, Johannes Gehrke, and Lars
  Vilhuber.
\newblock Privacy: Theory meets practice on the map.
\newblock In {\em IEEE 24th International Conference on Data Engineering
  (ICDE)}, pages 277--286, 04 2008.

\bibitem{hardt2010}
Moritz Hardt and Kunal Talwar.
\newblock On the geometry of differential privacy.
\newblock In {\em Proceedings of the Forty-Second ACM Symposium on Theory of
  Computing}, STOC '10, page 705–714, New York, NY, USA, 2010. Association
  for Computing Machinery.

\bibitem{ghosh2012}
Arpita Ghosh, Tim Roughgarden, and Mukund Sundararajan.
\newblock Universally utility-maximizing privacy mechanisms.
\newblock {\em SIAM Journal on Computing}, 41(6):1673--1693, 2012.

\bibitem{dwork2014}
Cynthia Dwork and Aaron Roth.
\newblock The algorithmic foundations of differential privacy.
\newblock {\em Foundations and Trends in Theoretical Computer Science},
  9(3–4):211--407, 2014.

\bibitem{dwork2016}
Cynthia Dwork and Guy~N. Rothblum.
\newblock Concentrated differential privacy, 2016.

\bibitem{rinott2018}
Yosef Rinott, Christine O'Keefe, Natalie Shlomo, and C.J. Skinner.
\newblock Confidentiality and differential privacy in the dissemination of
  frequency tables.
\newblock {\em Statistical Science}, 33:358--385, 08 2018.

\bibitem{fraser2005}
Bruce Fraser and Janice Wooton.
\newblock A proposed method for confidentialising tabular output to protect
  against differencing.
\newblock In {\em Monographs of Official Statistics: Work Session on
  Statistical Data Confidentiality}, pages 299--302, 11 2005.

\bibitem{marley2011}
Jennifer~K. Marley and Victoria~L. Leaver.
\newblock A method for confidentialising user-defined tables: Statistical
  properties and a risk-utility analysis.
\newblock In {\em Int. Statistical Inst.: Proc. 58th World Statistical Congress
  (Session IPS060)}, pages 1072--1081, 08 2011.

\bibitem{thompson2013}
Gwenda Thompson, Stephen Broadfoot, and Daniel Elazar.
\newblock Methodology for the automatic confidentialisation of statistical
  outputs from remote servers at the {Australian Bureau of Statistics}.
\newblock In {\em Joint UNECE/Eurostat work session on statistical data
  confidentiality}, 10 2013.

\bibitem{bailie2019}
James Bailie and Chien-Hung Chien.
\newblock {ABS} perturbation methodology through the lens of differential
  privacy.
\newblock In {\em Joint UNECE/Eurostat work session on statistical data
  confidentiality}, 10 2019.

\bibitem{abowd2018}
John~M. Abowd.
\newblock The {U.S. Census Bureau} adopts differential privacy.
\newblock In {\em Proceedings of the 24th ACM SIGKDD International Conference
  on Knowledge Discovery \& Data Mining}, KDD '18, page 2867, New York, NY,
  USA, 2018. Association for Computing Machinery.

\bibitem{garfinkel2019}
Simson~L. Garfinkel.
\newblock Deploying differential privacy for the 2020 census of population and
  housing.
\newblock In {\em JSM 2019 Session: Formal Privacy - Making an Impact at Large
  Organizations}, 07 2019.

\bibitem{petti2019}
Samantha Petti and Abraham Flaxman.
\newblock Differential privacy in the 2020 {US} census: what will it do?
  {Q}uantifying the accuracy/privacy tradeoff.
\newblock {\em Gates Open Research}, 3, 2019.

\bibitem{ruggles2019}
Steven Ruggles, Catherine Fitch, Diana Magnuson, and Jonathan Schroeder.
\newblock Differential privacy and census data: Implications for social and
  economic research.
\newblock {\em AEA Papers and Proceedings}, 109:403--08, May 2019.

\bibitem{santos2020}
Alexis~R. Santos-Lozada, Jeffrey~T. Howard, and Ashton~M. Verdery.
\newblock How differential privacy will affect our understanding of health
  disparities in the {United States}.
\newblock {\em Proceedings of the National Academy of Sciences},
  117(24):13405--13412, 2020.

\bibitem{essnet2017}
Laszlo Antal, Ma{\"e}l-Luc Buron, Annu Cabrera, Tobias Enderle, Sarah Giessing,
  Juno{\v s} Lukan, Eric Schulte~Nordholt, and Andreja Smukavec.
\newblock Harmonised protection of census data.
\newblock
  \url{https://ec.europa.eu/eurostat/cros/content/harmonised-protection-census-data_en},
  2017.
\newblock Accessed on 26 Aug 2020.

\bibitem{essnet2019}
Peter-Paul De~Wolf, Tobias Enderle, Alexander Kowarik, and Bernhard Meindl.
\newblock Perturbative confidentiality methods.
\newblock
  \url{https://ec.europa.eu/eurostat/cros/content/perturbative-confidentiality-methods_en},
  2019.
\newblock Accessed on 26 Aug 2020.

\bibitem{essnet2019a}
Peter-Paul De~Wolf, Tobias Enderle, Alexander Kowarik, and Bernhard Meindl.
\newblock {SDC Tools} - user support and sources of tools for statistical
  disclosure control.
\newblock \url{https://github.com/sdcTools}, 2019.
\newblock Accessed on 26 Aug 2020.

\bibitem{ashgar2019}
Hassan~Jameel Asghar and Dali Kaafar.
\newblock Averaging attacks on bounded noise-based disclosure control
  algorithms.
\newblock {\em Proceedings on Privacy Enhancing Technologies}, 2020(2):358 --
  378, 2020.

\bibitem{dwork2004}
Cynthia Dwork and Kobbi Nissim.
\newblock Privacy-preserving datamining on vertically partitioned databases.
\newblock In Matthew~K. Franklin, editor, {\em Advances in Cryptology - CRYPTO
  2004, 24th Annual International Cryptology Conference, Santa Barbara,
  California, USA, August 15-19, 2004, Proceedings.}, volume 3152 of {\em
  Lecture Notes in Computer Science}, pages 528--544. Springer, 2004.

\bibitem{blum2005}
Avrim Blum, Cynthia Dwork, Frank McSherry, and Kobbi Nissim.
\newblock Practical privacy: The {SuLQ} framework.
\newblock In {\em 24th ACM SIGMOD International Conference on Management of
  Data / Principles of Database Systems, Baltimore (PODS 2005)}, June 2005.

\bibitem{dwork2011}
Cynthia Dwork.
\newblock A firm foundation for private data analysis.
\newblock {\em Commun. ACM}, 54(1):86–95, January 2011.

\bibitem{dwork2010}
Cynthia Dwork and Moni Naor.
\newblock On the difficulties of disclosure prevention in statistical databases
  or the case for differential privacy.
\newblock {\em Journal of Privacy and Confidentiality}, 2(1), Sep. 2010.

\bibitem{bambauer2014}
J.R. Bambauer, Krish Muralidhar, and Rathindra Sarathy.
\newblock Fool's gold! {A}n illustrated critique of differential privacy.
\newblock {\em Vanderbilt J. Entertain. Technol. Law}, 16:701--755, 01 2014.

\bibitem{garfinkel2018}
Simson~L. Garfinkel, John~M. Abowd, and Christian Martindale.
\newblock Understanding database reconstruction attacks on public data.
\newblock {\em Queue}, 16(5):28–53, October 2018.

\bibitem{meindl2019}
Bernhard Meindl and Tobias Enderle.
\newblock {cellKey} - consistent perturbation of statistical tables.
\newblock In {\em Joint UNECE/Eurostat work session on statistical data
  confidentiality}, 10 2019.

\bibitem{giessing2016}
Sarah Giessing.
\newblock Computational issues in the design of transition probabilities and
  disclosure risk estimation for additive noise.
\newblock In Josep Domingo{-}Ferrer and Mirjana Pejic{-}Bach, editors, {\em
  Privacy in Statistical Databases - {UNESCO} Chair in Data Privacy,
  International Conference, {PSD} 2016, Dubrovnik, Croatia, September 14-16,
  2016, Proceedings}, volume 9867 of {\em Lecture Notes in Computer Science},
  pages 237--251. Springer, 2016.

\bibitem{enderle2020}
Tobias Enderle, Sarah Giessing, and Reinhard Tent.
\newblock Calculation of risk probabilities for the cell key method.
\newblock In Josep Domingo{-}Ferrer and Krishnamurty Muralidhar, editors, {\em
  Privacy in Statistical Databases - {UNESCO} Chair in Data Privacy,
  International Conference, {PSD} 2020, Tarragona, Spain, September 23-25,
  2020, Proceedings}, volume 12276 of {\em Lecture Notes in Computer Science},
  pages 151--165. Springer, 2020.

\bibitem{enderle2018}
Tobias Enderle, Sarah Giessing, and Reinhard Tent.
\newblock Designing confidentiality on the fly methodology - three aspects.
\newblock In Josep Domingo{-}Ferrer and Francisco Montes, editors, {\em Privacy
  in Statistical Databases - {UNESCO} Chair in Data Privacy, International
  Conference, {PSD} 2018, Valencia, Spain, September 26-28, 2018, Proceedings},
  volume 11126 of {\em Lecture Notes in Computer Science}, pages 28--42.
  Springer, 2018.

\bibitem{mervis2019}
Jeffrey Mervis.
\newblock Can a set of equations keep us census data private?
\newblock {\em Science}, 10, 2019.

\bibitem{mervis2018}
Jeffrey Mervis.
\newblock Trump officials claim they can avoid 2020 census problems caused by
  controversial citizenship question. experts are very skeptical.
\newblock {\em Science}, 04, 2018.

\bibitem{hansen2018}
Mark Hansen.
\newblock To reduce privacy risks, the census plans to report less accurate
  data.
\newblock \url{https://nyti.ms/2E4UeZQ}, 2018.
\newblock Accessed on 28 Aug 2020.

\bibitem{dwork2010a}
Cynthia Dwork and Adam Smith.
\newblock Differential privacy for statistics: What we know and what we want to
  learn.
\newblock {\em Journal of Privacy and Confidentiality}, 1(2), Apr. 2010.

\bibitem{wang2015}
Yue Wang, Jaewoo Lee, and Daniel Kifer.
\newblock Revisiting differentially private hypothesis tests for categorical
  data, 2015.

\bibitem{bach2018}
Fabian Bach.
\newblock Statistical disclosure control in geospatial data: The 2021 {EU}
  census example.
\newblock In J{\"u}rgen D{\"o}llner, Markus Jobst, and Peter Schmitz, editors,
  {\em Service-Oriented Mapping: Changing Paradigm in Map Production and
  Geoinformation Management}, pages 365--384. Springer International
  Publishing, Cham, 2019.

\bibitem{costemalle2019}
Vianney Costemalle.
\newblock Detecting geographical differencing problems in the context of
  spatial data dissemination.
\newblock {\em Statistical Journal of the IAOS}, 35(4):559–568, Dec. 2019.

\end{thebibliography}
\end{document}